\def\PR{Phys. Rev.\ }

\def\JPCM{J. Phys. Condens. Matter\ }

\def\etal{{\it et.al.}}
\def\r{{\mathbf r}_i}
\def\k{{\mathbf k}\ }
\def\n{{\noindent}}

\def\g{{\mathbf g}}

\def\A{{\mathbf A}}
\def\B{{\mathbf B}}
\def\F{{\mathbf F}}
\def\S{{\mathbf S}}
\def\mbf{\mathbf }
\def\pls{\enskip +\enskip}
\def\mns{\enskip -\enskip}
\def\wt{\widetilde }
\documentclass[twocolumn,showpacs,preprintnumbers,amsmath,amssymb]{revtex4}
\usepackage{graphicx}% Include figure files
\usepackage{dcolumn}% Align table columns on decimal point
\usepackage{bm}% bold math
\begin{document}
\title [Lattice thermal conductivity of disordered binary alloys]{Lattice thermal conductivity of disordered binary alloys I : a formulation}
\author{Aftab Alam}
\email{alam@bose.res.in}
\author{Abhijit Mookerjee}
\email{abhijit@bose.res.in}
\affiliation{S.N. Bose National Centre for Basic Sciences,
 JD Block, Sector III, Salt Lake City, Kolkata 700098,
India}
\begin{abstract}
We present here a formulation for the calculation of the configuration averaged lattice thermal conductivity in random alloys. Our formulation is based on the augmented-space theorem, introduced by one of us \cite{am}, combined with a generalized diagrammatic technique. The diagrammatic approach simplifies the problem of including effects of disorder corrections to a great extent. The approach allows us to obtain an expression for the effective heat current in case of disordered alloys, which in turn is used in a Kubo-Greenwood type formula for the thermal conductivity. We show that disorder scattering renormalizes the phonon propagators as well as the heat currents. The corrections to the current terms have been shown to be related to the self-energy of the propagators. We also study the effect of vertex corrections in a simplified ladder diagram approximation. A mode dependent diffusivity $D_{\gamma}$ and then a total thermal diffusivity averaged over different modes are defined. Schemes for implementing the said formalism are discussed. A few initial numerical results on the frequency and temperature dependence of lattice thermal conductivity are presented for NiPd alloy and are also compared with experiment. We also display numerical results on the frequency dependence of thermal diffusivity averaged over modes.
\end{abstract}

\maketitle
\section{Introduction}
The problem of phonon excitations have been studied extensively both theoretically and experimentally in mixed crystals \cite{elliot,kambrock} as well as in insulators \cite{maradudin}. However, there
are far fewer studies of lattice thermal conductivity in disordered alloys. Detailed comparison
between theory and experiment on the basis of realistic models has not been extensive. Model
calculations are mostly based on mass disorder, whereas in phonon problems essential off-diagonal disorder in the force constants cannot be dealt within single site mean-field approximations. 
Such disorder cannot be ignored in a realistic calculation.
 
In the past few decades, there has been considerable attention directed towards the theoretical understanding of the lattice thermal conductivity of metals. These are mainly due to the efforts of Klemens {\cite{klemens}}, Ziman {\cite{ziman}}, Callaway {\cite{call}}, Parrot {\cite{parrot}} and others.
 However majority of these were based on model calculations, either for perfect crystals or ordered alloys.
 Flicker and Leath \cite{FL} first proposed the calculation of lattice thermal conductivity within a  single-site coherent potential approximation (CPA) using the appropriate Kubo formula. The single-site CPA is a  mean field approximation, capable of dealing with mass disorder alone and is not adequate for treating intrinsic off-diagonal disorder arising out of the force constants.  This was evidenced in the inability of the single site CPA to explain experimental life time data on NiPt \cite{Tsunoda}.
 Translationally invariant, multiple site,  multiple scattering theories based on the augmented space formalism\ \cite{am} have recently been proposed  by  Ghosh \etal\ \cite{glc} as well as by us \cite{am1}
 to describe  phonons in a series of random alloy systems :  NiPt, NiPd and NiCr. These formalisms explicitly capture the effects of both the diagonal and off-diagonal disorder.

In this paper, we shall first introduce a Kubo-Greenwood type formula which relates the thermal conductivity to the (heat) current-current correlation function. The ideas used here are  very similar to those proposed by Allen and Feldman \cite{af} except that the present formulation is done keeping in mind the application to a substitutionally disordered crystal, rather than an amorphous system. For disordered alloys, configuration averaging over various random atomic arrangements have been carried out using the augmented space formalism (ASF) introduced by us \cite{am}. The ASF goes beyond the usual mean-field approaches and takes into account configuration fluctuations over a large local environment. We shall combine the augmented space representation for phonons \cite{am1} with a scattering diagrammatic  technique to get an effective heat current. This effective current consists in addition to the averaged current term, also the terms arising out of the disorder scattering corrections. We will show that these disorder induced corrections to the averaged current terms are directly related either to the disorder scattering induced self-energy matrix in the propagator or to vertex corrections. As far as the vertex corrections are concerned, Leath \cite{leath} had obtained these corrections within the framework of CPA by using diagram summations. In this paper we shall derive the contribution of these corrections in a more generalized context with the inclusion of diagonal as well as the intrinsic off-diagonal disorder arising out of the dynamical matrix.  Since in an earlier communication \cite{am2} we have already shown that the self energy matrix and the Green matrix  can be calculated for realistic binary alloys within an augmented space block recursion (ASBR) technique, so the present formulation will form the basis of a subsequent calculation of lattice thermal conductivity in realistic alloys.

The rest of the paper is organized as follows. In Sec. II, we describe the basic tools used to calculate the lattice thermal conductivity for a crystal. In Sec. III(A), we briefly introduce the augmented space representation for phonons. In Sec. III(B,C), we derive expressions for important physical quantities such as effective heat current, averaged lattice thermal conductivity and thermal diffusivity in terms of configuration averaged Green matrix and self energy matrix of the system. In Sec. III(D), we describe vertex corrections arising out of the correlated propagation. Sec. IV(A,B) is devoted for a description of the schemes for implementing this formalism to realistic alloys. Few of the initial numerical results on the lattice thermal conductivity and thermal diffusivity for NiPd alloy are shown in Sec. V and an effort has been put forward to compare them with the available experimental data.  Concluding remarks appear in Sec. VI.
\section{Thermal  conductivity}

In an earlier work \cite{sm} we had reported a formulation for the study of optical conductivity
in disordered alloys. Here we attempt to modify that formulation for lattice thermal conductivity studies also
in disordered alloys. The formulations have similar overall similarities, but differ in specific
details, which we would like to focus on in this communication.
The Kubo formula which relates the optical conductivity to a current-current correlation function
is well established. The Hamiltonian contains a term $\sum_i {\mbf j}_i\cdot{\mbf A}({\r},t)$ which drives the electrical current. For thermal conductivity we do not have a similar term in the Hamiltonian which drives a heat current. The derivation of a Kubo formula in this situation requires an additional statistical hypothesis \cite{af}, which states that a system in steady state has a space dependent {\sl local} temperature $T({\r}) = [\kappa_B \beta({\r})]^{-1}$. 
The expression for the heat current has been discussed in great detail by Hardy \cite{hardy} and 
Allen and Feldman \cite{af}. The readers are refereed to these papers
for the details of calculation. The matrix element of the heat current in the basis of the eigenfunctions of the Hamiltonian is given by~:

\begin{equation}
{\mbf S}_{\gamma\gamma^\prime}^\mu (\k)\ =\ \frac{\hbar}{2}\ \left(\rule{0mm}{3mm} \omega_{\k\gamma}+\omega_{\k\gamma^\prime}\right)\ {\mbf v}^\mu_{\gamma\gamma^\prime}(\k) ,
\end{equation}

\n where, the phonon group velocity ${\mbf v}_{\gamma\gamma^\prime}(\k)$ is given by

\begin{eqnarray}
{\mbf v}_{\gamma\gamma^\prime} & = & \frac{i}{2\sqrt{\omega_{\k\gamma}\omega_{\k\gamma^\prime}}}\ 
\sum_\mu\sum_\nu \epsilon^\mu_\gamma(\k)\ \epsilon^\nu_{\gamma^\prime}(\k)\nonumber\\
& &  \left( \sum_{\mbf R_{ij}}\ \frac{\Phi^{\mu\nu}({\mbf R_{ij}})}
{\sqrt{M_iM_j}}\ \right){\mbf R}_{ij}\  e^{i\k\cdot{\mbf R}_{ij}}\nonumber\\ 
 & = & \frac{1}{2\sqrt{\omega_{\k\gamma}\omega_{\k\gamma^\prime}}}\                                 
\sum_\mu\sum_\nu \epsilon^\mu_\gamma(\k)\ {\mbf \nabla}_\k D^{\mu\nu}(\k) \ \epsilon^\nu_{\gamma^\prime}(\k)\nonumber ,\\
\end{eqnarray}

\n here $\gamma,\gamma^\prime$ label the various modes of vibration, $\omega_{\k\gamma},\omega_{\k\gamma^\prime}$ are their frequencies,  $\epsilon^\mu_\gamma(\k), \epsilon^\nu_{\gamma^\prime}(\k)$ are the polarization vectors and $D^{\mu\nu}(\k)$ is the Fourier transform of mass scaled dynamical matrix.

We shall consider the case where the temperature gradient is uniform within the system. 
The Kubo formula then relates the linear heat current response to the temperature gradient  field 

\[ \langle S^{\mu}(t) \rangle = - \sum_{\nu}\int_{-\infty}^{\infty} \ dt'\ \kappa^{\mu\nu}(t-t')\  
{\mbf \nabla}^\nu \delta T(t),  \]
\noindent where

\[ 
\kappa^{\mu\nu}(\tau) =  \Theta(\tau)\ \frac{1}{T}\int_0^\beta\  d\lambda \langle  S^{\mu}(-i\hbar\lambda),S^{\nu}(\tau) \rangle ,
\]

$\Theta(\tau)$ is the Heaviside step function,
and
\[
S(-i\hbar\lambda)\ =\ e^{\lambda H}\ S\ e^{-\lambda H}.
\]

\n $\langle\ \rangle$ on the right-hand side of the above equation denotes thermal averaging over 
states in the absence of the temperature gradient. The above equation can be rewritten in the form of a Kubo-Greenwood expression 
\begin{eqnarray}
\kappa^{\mu\nu}(\omega, T) & =& \ \kappa^{\mu\nu}_I(\omega, T) \ +\ \kappa^{\mu\nu}_{II}(\omega, T) \nonumber\\
\phantom{x} \nonumber\\
 \kappa^{\mu\nu}_I (\omega, T)& =& \frac{\pi}{T}\ \int \frac{d^3\k}{8\pi^3}\ \sum_\gamma\sum_{\gamma^\prime\ne\gamma}\ \frac{\langle n_{\k\gamma^\prime}\rangle-\langle n_{\k\gamma}\rangle}{\hbar(\omega_{\k\gamma}-\omega_{\k\gamma^\prime})}\nonumber\\ 
&&{\mbf S}^\mu_{\gamma\gamma^\prime}(\k){\mbf S}^\nu_{\gamma^\prime\gamma}(\k) \ \delta(\omega_{\k\gamma}-\omega_{\k\gamma^\prime}-\omega) \nonumber\\
\phantom{x}\\
 \kappa^{\mu\nu}_{II}(\omega, T)& = &\frac{1}{\kappa_B T^2} \left[ \rule{0mm}{4mm}\left\{ \int\frac{d^3\k}{8\pi^3} \sum_\gamma \langle n_{\k\gamma}\rangle\ {\mbf S}^\mu_{\gamma\gamma}(\k)\right\}\right.\nonumber\\ 
&&\left. \left\{ \int\frac{d^3\k}{8\pi^3}\sum_\gamma\langle n_{\k\gamma}\rangle\ {\mbf S}^\nu_{\gamma\gamma}(\k)\right\} - \kappa_B T \int\frac{d^3\k}{8\pi^3}\right.\nonumber\\
&& \left.\sum_\gamma \frac{\partial\langle n_{\k\gamma}\rangle }
{\partial(\hbar\omega_{\k\gamma})}\ S^\mu_{\gamma\gamma}(\k)\ S^\nu_{\gamma\gamma}(\k) \right]\delta(\omega) ,
\end{eqnarray}
where $\langle n_{\k\gamma}\rangle = (e^{\beta\hbar\omega_{\k\gamma}}-1)^{-1}$ is the equilibrium Bose Einstein distribution function and T is the absolute temperature.

The first expression is for inter-band transitions, while the second expression is for intra-band transitions.
For an isotropic response, we can rewrite the first expression as 

\begin{eqnarray*}
 \kappa_I(\omega, T)&=& \frac{\pi}{3T}\sum_{\mu} \int\ d\omega^\prime \int\frac{d^3\k}{8\pi^3} \sum_{\gamma}\sum_{\gamma^{\prime}}
\widehat{\mbf S}^\mu_{\gamma\gamma^\prime}(\k, T) \nonumber\\
&& \widehat{\mbf S}^\mu_{\gamma^{\prime} \gamma}(\k, T) \delta(\omega^\prime - \omega_{\k\gamma^\prime})\delta(\omega^\prime+\omega-\omega_{\k\gamma}) ,
\end{eqnarray*}

\n where 

\[
\widehat{\mbf S}^\mu_{\gamma\gamma^\prime}(\k,T)\ =\ \sqrt{\left|\frac{\langle n_{\k\gamma^\prime}\rangle-\langle n_{\k\gamma}\rangle}{\hbar(\omega_{\k\gamma}-\omega_{\k\gamma^\prime})}\right|}\ {\mbf S}^\mu_{\gamma\gamma^\prime}(\k).
\]

We may rewrite the above equation as 

\begin{eqnarray*}
\kappa_I(\omega,T) &=& \frac{1}{3\pi T}\ \sum_{\mu}\ \int d\omega^\prime\int\frac{d^3\k}{8\pi^3} \mbox{Tr} \left[\rule{0mm}{4mm}\ \widehat{\mbf S}^{\mu}(\k,T)\right.\nonumber\\ &&\left.\hspace{-8.0mm} \Im m \{ {\mbf G}(\k,\omega^\prime)\}\ \widehat{\mbf S}^{\mu}(\k,T)\ \Im m\{ {\mbf G}(\k,\omega^\prime+\omega)\} \right].
\end{eqnarray*}

The operator {\bf G}($\omega$) is the phonon Green operator $(M\omega^2{\mbf I}-{\mbf\Phi})^{-1}$. The Trace is invariant in different representations. For crystalline systems, usually the Bloch basis 
$\{\vert {\mbf k},\gamma\rangle\}$ is used. For disordered systems, prior to configuration averaging,
  it is more convenient to use the basis $\{\vert \k,\alpha\rangle\}$, where $\k$ is the reciprocal vector and $\alpha$
represents the coordinate axes directions.  We can transform from the mode basis to the coordinate basis by
using the transformation matrices $\Upsilon_{\gamma\alpha}(\k) =\ \epsilon^\alpha_\gamma(\k)$. For example 

\[ \widehat{\mbf S}_{\alpha\beta}^\mu(\k,T) \ =\     \Upsilon^{-T}_{\alpha\gamma}(\k)\ \widehat{\mbf S}^\mu_{\gamma\gamma^\prime}(\k,T)\ \Upsilon^{-1}_{\gamma^{\prime}\beta}(\k).
\]

\noindent If we define

\begin{equation} 
{\mbf \kappa}(z_1,z_2)  = \int\frac{d^3\k}{8\pi^3}\ \mbox{Tr} \left[\ \rule{0mm}{4mm}\widehat{\mbf S}\  {\mbf G}(\k,z_1)\ \widehat{\mbf S}\ {\mbf G}(\k,z_2)\right] .
\label{ref1}
\end{equation}

\noindent  then the above equation becomes,
\begin{eqnarray}
 \kappa_I(\omega,T) &=&  \frac{1}{12\pi T}\ \sum_{\mu}  \int d\omega^\prime \ \left[\rule{0mm}{4mm} {\mbf \kappa}^{\mu\mu}(\omega^{\prime -},\omega^{\prime +}+\omega)\right.\nonumber \\ &&\left. + {\mbf \kappa}^{\mu\mu}(\omega^{\prime +},\omega^{\prime -}+\omega)   - {\mbf \kappa}^{\mu\mu}(\omega^{\prime +},\omega^{\prime  +}+\omega)\right.\nonumber\\ && \left. - {\mbf \kappa}^{\mu\mu}(\omega^{\prime -},\omega^{\prime  -}+\omega)\right]\rule{0mm}{4mm},
\label{ref2}
\end{eqnarray}

\noindent where

\[ 
f(\omega^+) = \lim_{\delta\rightarrow 0} f(\omega+i\delta),\phantom{xx}
f(\omega^-) = \lim_{\delta\rightarrow 0} f(\omega-i\delta).
\]

We have used the herglotz analytic property \cite{am1} of the Green operator   

\[
{\mbf G}(\omega+i\delta) = \Re e\left[\rule{0mm}{3mm}{\mbf G}(\omega)\right] \mns  i\ \mbox{sgn}(\delta)\ \Im m\left[\rule{0mm}{3mm}{\mbf G}(\omega)\right]. 
\]

For disordered materials, we shall be interested in obtaining the configuration averaged  response functions. This will require
the configuration averaging of quantities like $\kappa(z_1,z_2)$

We should note that the expressions (\ref{ref1}) and (\ref{ref2}) are similar to the corresponding equations for
optical conductivity, with the heat current replacing electrical current. To deal with random alloys we shall follow an analogous procedure for optical conductivity proposed earlier \cite{sm}

\section{Configuration averaging }
\subsection{The augmented space formalism for  phonons}

The augmented space formalism (ASF) has been described in detail in several earlier papers (see \cite{mook2}-\cite{mook3}). We shall, for the sake
of completeness,  describe only those features which will be necessary for the implementation of our ideas in this
communication. It is also important to introduce the notations used subsequently. 
The theory of phonons consists of solving a secular equation of the form 
 \[ ({\bf M}w^{2} - {\bf D})\ {\bf u}(R,w) = 0, \]  where $u_{\alpha}(R,w)$ is the Fourier transform of $u_{\alpha}(R,t)$, the displacement of an atom from its equilibrium position $R$ on the lattice, in the direction ${\alpha} $ at time $t$. {\bf M} is the {\it mass operator}, diagonal in real-space
 and {\bf D} is the {\it dynamical matrix operator} whose tight-binding representations are
\begin{eqnarray}
{\bf M} &=& \sum_{R}  m_{R}\ {\delta}_{\alpha \beta} \ P_R, \nonumber\\
{\bf D} &=&  \sum_{R} \left\{\sum_{R^{\prime} \ne R}\Phi_{RR^{\prime}}^{\alpha \beta}\right\}\ P_{R} + \sum_{R}\sum_{R^{\prime} \ne R} \Phi_{RR^{\prime}}^{\alpha \beta}\ T_{RR^{\prime}},\nonumber\\
\end{eqnarray}
\n where the sum rule has been incorporated in the first term of the equation involving $\bf D$. \\
Here $P_R$ is the projection operator\ $\vert R\rangle\langle R\vert$\ \ and $T_{RR'}$ is the transfer operator\ $\vert R\rangle\langle R'\vert$\ \ in the Hilbert space ${\cal H}$ spanned by the tight-binding basis $\{\vert R\rangle\}$.
 $R,R^{\prime}$ specify the lattice sites and $\alpha $,$ \beta $ the Cartesian directions. $m_{R}$ is the mass of an atom occupying the position $R$ and $\Phi_{RR^{\prime}}^{\alpha \beta}$ is the force constant tensor. 

\noindent We shall be interested in calculating the displacement-displacement Green matrix ${\mbf G} (R,R',w^2)$ 

\[ {\mbf G}(R,R',w^2) = \langle R | \left({\bf{M}}w^{2}-{\bf{D}}\right)^{-1} | R' \rangle . \]

Let us now consider a binary alloy $ A_{x}B_{y} $ consisting of two kinds of atoms A and B of masses
 $m_A$ and  $m_B$ randomly occupying each lattice sites. We wish to calculate the configuration-averaged
 Green matrix $\ll {\bf G}({R,R'},w^2)\gg$. We shall use the augmented space formalism to do so indicating the main operational results here.  For further details we  refer the reader to the  monograph \cite{tf}.
The first operation is to represent the random parts of the secular equation in terms of a random
set of local variables $\{ n_R\}$ which are 1 if the site $R$ is occupied by an  A  atom and 0
if it is occupied by B. The probability densities of these variables may be written as
 
\begin{eqnarray}
 Pr(n_R)& = & x\ \delta (n_{R}-1)\ +\ y\ \delta(n_{R})\nonumber\\
   & =  & (-1/{\pi})\ \Im m\langle {\uparrow}_{R} |\ (n_{R}I-{\it {N_{R}}})^{-1}\ | {\uparrow}_{R}\rangle , \label{prob}
\end{eqnarray}

\noindent where $x$ and $y$ are the concentrations of the constituents A and B with $x+y=1$. $N_{R}$ is an operator defined on the configuration-space $\phi_{R}$ of the variable $n_{R}$. This is of rank $2$ and is spanned by the states $\{|{\uparrow_{R}}\rangle, |{\downarrow_{R}}\rangle \}$,
\[ \vert\uparrow_R\rangle\ =\ \sqrt{x}\vert A_R\rangle + \sqrt{y}\vert B_R\rangle, \quad\vert\downarrow_R\rangle\ =\ \sqrt{y}\vert A_R\rangle - \sqrt{x}\vert B_R\rangle,\] 
\[ N_{R} = xp_{R}^{\uparrow} + yp_{R}^{\downarrow} + \sqrt{xy}\ {\cal T}^{\uparrow\downarrow}_R, \quad\quad{\cal T}^{\uparrow\downarrow}_R\ =\ {\tau}^{\uparrow \downarrow}_{R} + {\tau}^{\downarrow \uparrow}_{R}, \]

\noindent where $\vert A_R\rangle$ is the state in which an atom of the type A occupies a site $R$. $\vert B_R\rangle$ is similarly defined. 
 $p_R^\uparrow =\vert\uparrow_R\rangle\langle\uparrow_R\vert$ and $p_R^\downarrow=\vert\downarrow_R\rangle\langle\downarrow_R\vert$ are projection operators 
and  ${\tau}_R^{\uparrow\downarrow}=\vert\uparrow_R\rangle\langle\downarrow_R\vert$ and
 ${\tau}_R^{\downarrow\uparrow}=\vert\downarrow_R\rangle\langle\uparrow_R\vert$ are transfer operators in the configuration space $\phi_{R}$. 

In terms of random variables $n_{R}$, the mass operator can be written as 
\begin{equation} 
{\bf M}\ =\ \sum_{R}  \left[\rule{0mm}{4mm} m_{B}\ +\ n_{R}\ (\delta m) \right] \delta_{\alpha \beta} \ P_R,  \quad \quad \delta m=m_A-m_B
\end{equation}

\noindent According to the augmented space theorem, in order to obtain the configuration-average we simply replace the random variables $n_R$ by the
corresponding operators $N_R$ associated with its probability density, as in Eqn. (\ref{prob}), and take the matrix element of
the resulting operator between the {\sl reference states}. For a full
mathematical proof the reader is referred to \cite{tf}.

\[
 n_{R}\longrightarrow N_{R} \ =\  
x\ \tilde{I}\ +\ (y-x)\  p_{R}^{\downarrow} + \sqrt{xy}\ {\cal T}^{\uparrow \downarrow}_R .  
\]

\noindent Using the above we get,

%\begin{widetext}
\begin{eqnarray}
\widetilde{\bf M}&=& \A({\mbf m} )\ \widetilde{I}\otimes  I + \B({\mbf m} )\ \sum_{R} p_{R}^\downarrow \otimes P_{R} \nonumber \\ & &+ \F({\mathbf m} )\ \sum_{R}\  {\cal T}^{\uparrow\downarrow}_R \otimes P_{R}\nonumber\\ 
\phantom{\widetilde{\bf M}} &=& \ll \widetilde{\mbf M}\gg\  +\  \widetilde{\mbf M}^\prime,
\label{mass}
\end{eqnarray}
%\end{widetext}
\noindent where

\[   \begin{array}{ll}

\A({\mathbf X}) = \ll {\mbf X} \gg\ =(x {\mathbf X}_A+y {\mathbf X}_B), \\
\B({\mathbf X}) = (y-x)\ ({\mathbf X}_A-{\mathbf X}_B), \\
\F({\mathbf X}) = \sqrt{xy} \ ({\mathbf X}_A-{\mathbf X}_B).  \end{array}  \]

\noindent Similarly the random off-diagonal force constants $\Phi_{RR^{\prime}}^{\alpha \beta}$ between the sites $R$ and $R^{\prime}$ can be written as 

\begin{eqnarray}
\Phi_{RR^{\prime}}^{\alpha \beta} &=& \Phi_{AA}^{\alpha \beta} n_{R} n_{R^{\prime}} + \Phi_{BB}^{\alpha \beta} (1-n_{R}) (1-n_{R^{\prime}}) +\nonumber\\
&&  \Phi_{AB}^{\alpha \beta} \left[\rule{0mm}{4mm}\ n_{R}(1-n_{R^{\prime}}) + n_{R^{\prime}}(1-n_{R})\ \right]\nonumber\\
\phantom{x} \nonumber\\
&=&  \Phi_{BB}^{\alpha \beta}\ +\ \left(\rule{0mm}{4mm}\Phi_{AA}^{\alpha \beta} + \Phi_{BB}^{\alpha \beta} - 2 \Phi_{AB}^{\alpha \beta}\right)\ n_{R} n_{R^{\prime}} +\nonumber\\
&&\left(\rule{0mm}{4mm}\Phi_{AB}^{\alpha \beta} - \Phi_{BB}^{\alpha \beta}\right)\ (n_{R} + n_{R^{\prime}}). \nonumber \\
\end{eqnarray}

\noindent Let us define the following :

\begin{eqnarray*}
\Phi^{\alpha\beta}_{(1)} &=& x\ \Phi_{AA}^{\alpha \beta} - y\ \Phi_{BB}^{\alpha \beta} + (y-x) \Phi_{AB}^{\alpha\beta},\\
\Phi^{\alpha\beta}_{(2)} &=& \Phi_{AA}^{\alpha \beta} + \Phi_{BB}^{\alpha \beta} - 2 \Phi_{AB}^{\alpha \beta}.
\end{eqnarray*}

\noindent In augmented space the off-diagonal force constant matrix becomes an operator 

\begin{eqnarray*}
\widetilde{\mbf D}^{\alpha\beta}_{(off)} &=& \sum_{RR'}\ \left[\rule{0mm}{5mm} \ll \Phi^{\alpha\beta}_{RR'}\gg\ \tilde{I} + 
  \Phi^{\alpha\beta}_{(1)} \left\{ (y-x)\ (p^\downarrow_R
+p^\downarrow_{R'})\right.\right.\nonumber\\
&& \left.\left.\hspace{-0.5in} +\sqrt{xy} ({\cal T}^{\uparrow\downarrow}_{R}+{\cal T}^{\uparrow\downarrow}_{R'})\right\}+ \Phi^{\alpha\beta}_{(2)}\ \left\{ (y-x)^2\ p^\downarrow_R\ p^\downarrow_{R'}+ \right.\right.\\ 
&&\left.\left. \hspace{-0.45in}\sqrt{xy}(y-x) \left(p^\downarrow_R {\cal T}^{\uparrow\downarrow}_{R'} + p^\downarrow_{R'} {\cal T}^{\uparrow\downarrow}_{R}\right) + xy {\cal T}^{\uparrow\downarrow}_{R}{\cal T}^{\uparrow\downarrow}_{R'} \right\}\rule{0mm}{5mm} \right]\otimes T_{RR'} \\
\phantom{x}\\
&=& \sum_{RR'} \ll \Phi^{\alpha\beta}_{RR'}\gg \tilde{I}\otimes T_{RR'} +  \sum_{RR'}\ \Psi_{RR'}^{\alpha\beta}\otimes T_{RR'}. \\
\end{eqnarray*}

\noindent The sum rule gives the diagonal element,
\begin{eqnarray*}
\widetilde{\mbf D}^{\alpha\beta}_{(dia)}& =& -\sum_{R}\left\{\rule{0mm}{1mm}\sum_{R'\ne R} \ll \Phi^{\alpha\beta}_{RR'}\gg \widetilde {I}\right\} \otimes P_R \\
& &  - \sum_{R}\ \left\{ \sum_{R'\ne R}  \Psi_{RR'}^{\alpha\beta}\right\}\otimes P_R
\end{eqnarray*}

\noindent The total dynamical matrix in the augmented space is 

\begin{eqnarray}
 \widetilde{\mathbf D} &\ =\ & \ll\widetilde{\mbf D}\gg  - \sum_{R}\ \left\{ \sum_{R'\ne R} \Psi_{RR'}^{\alpha\beta}\right\}\otimes P_R \ \nonumber\\
& &  \ +\ 
 \sum_{RR'}\ \Psi_{RR'}^{\alpha\beta}\otimes T_{RR'} \nonumber\\
 \phantom{\widetilde{\mathbf D}} &\ =\ & \ll\widetilde{\mbf D}\gg\ +\ \widetilde{\mbf D}^\prime.  
\label{dm}
\end{eqnarray}

\n The boldface operators are $3\times 3$ matrix representations in the three Cartesian directions.

The augmented space theorem \cite{am} now states that the configuration-average of the Green matrix $\ll~{\bf G}({R,R'},w^2)~\gg$ may be written as 
\begin{eqnarray}
\ll {\mbf G}\left({R,R'},w^{2}\right)\gg \phantom{xxxxxxxxxxxxxxxxxxxxxx}\nonumber\\
\phantom{xxx}  =  \langle \{ \emptyset \}\otimes R|\ \left(\widetilde{\bf {M}}\  w^{2} -  \widetilde{\bf {D}}\right)^{-1}\   |\{\emptyset\} \otimes R'\rangle\ ,\phantom{xx}
\label{g0}
\end{eqnarray}
\noindent where $\widetilde{\bf{M}}$ and $\widetilde{\bf {D}}$ are the operators 
which are constructed out of ${\bf{M}}$
 and ${\bf {D}}$ by replacing all the random variables $n_{R}$ (or $n_{R^{\prime}})$ by 
the corresponding operators $N_{R}$ (or $N_{R^{\prime}})$ as given by Eqn.(\ref{mass}) and (\ref{dm}). These are the operators in the augmented space $ \Omega = {\cal H} \otimes \Phi $. The state $|{R} \otimes \{\emptyset\}\rangle$ is 
a state in the  augmented space, which is the direct product of the real-space and the configuration-space bases.The configuration-space $ \Phi = \prod_{R}^{\otimes}\phi_{R} $ is of rank $2^{N}$ for a system of N-lattice sites with binary distribution. A basis in this space is denoted by the cardinality sequence $ \{{\cal C}\} = \{R_{1},R_{2},\ldots,R_{c}\} $  which gives us the positions where we have a $\vert\!\downarrow \rangle$ configuration. The configuration $\{\emptyset\}$ refers to a null cardinality sequence i.e. one in which we have $\vert\uparrow \rangle$ at all sites.

The {\sl virtual crystal } (VCA) Green matrix is 
\begin{equation}
 {\mbf g} ({R,R'},w^{2}) = \langle  \{\emptyset\}\otimes R \vert ( \ll \widetilde{\mbf M}\gg \omega^{2}  - \ll\widetilde{\bf D}\gg)^{-1} \vert \{\emptyset\} \otimes R'\rangle ,
\label{vcgreen}
\end{equation} 
\n where
\[ \ll\widetilde{\mbf M}\gg = \ll m\gg \widetilde{I}\otimes I . \]
\noindent Referring back to Equations (\ref{mass}),(\ref{dm}) and (\ref{g0}) we get 

\begin{eqnarray}
\ll {\mbf G}(R,R',w^2)\gg &=& 
 \langle  \{\emptyset\}\otimes R \vert\ \left( \rule{0mm}{4mm} \ll \widetilde{\mbf M}\gg \omega^{2}- \ll\widetilde{\bf D}\gg\right.\nonumber\\
&&\left.  +\ \widetilde{\mbf M}'\omega^2 - \widetilde{\mbf D}'\right)^{-1}\ \vert \{\emptyset\} \otimes R'\rangle \nonumber\\
&=& \langle\{\emptyset\}\otimes R\vert \left({\mbf g}^{-1}- \widetilde{\mbf D}_1\right)^{-1}\vert\{\emptyset\}\otimes R'\rangle, \nonumber
\end{eqnarray}
\begin{equation}
\label{main}
\end{equation}
\n we define 
\begin{eqnarray*}
\widetilde{\mathbf D}_1 &=& \sum_R \  \left\{\rule{0mm}{5mm} {- \mbf \Upsilon}_R -\sum_{R'\ne R} {\mathbf\Psi}_{RR'}\right\}\ \otimes P_R +\nonumber\\ 
&&\sum_{R}\sum_{R'\ne R}\  {\mathbf\Psi}_{RR'}\otimes T_{RR'}
\end{eqnarray*}
\n with
\begin{eqnarray}
{\mbf\Upsilon}_R & = & \B({\mbf m})\ w^2\ p^\downarrow_R \ +\ \F({\mbf m})\ w^2\ {\cal T}^{\uparrow\downarrow}_R
\nonumber\\
\phantom{x}\nonumber\\
 {\mathbf\Psi}_{RR'}& = & {\mathbf D}^{(1)}_{RR'}\  \left(p^\downarrow_R +  p^\downarrow_{R'}\right)
+ {\mathbf D}^{(2)}_{RR'}\ \left({\cal T}^{\uparrow\downarrow}_R + {\cal T}^{\uparrow\downarrow}_{R'}\right) +\nonumber\\
&& {\mathbf D}^{(3)}_{RR'}\  p^\downarrow_R\  p^\downarrow_{R'}+ {\mathbf D}^{(4)}_{RR'}\ \left( p^\downarrow_R\ {\cal T}^{\uparrow\downarrow}_{R'} + 
 {\cal T}^{\uparrow\downarrow}_{R}\ p^\downarrow_{R'}\right)\nonumber\\
&& +\ {\mathbf D}^{(5)}_{RR'}\ {\cal T}^{\uparrow\downarrow}_R
\ {\cal T}^{\uparrow\downarrow}_{R'},
\label{dyn}
\end{eqnarray}
\n where
\begin{eqnarray*}
{\mathbf D}^{(1)}&=&(y-x)\ \Phi_{(1)},\\
{\mathbf D}^{(2)}&=&\sqrt{xy}\ \Phi_{(1)},\\
{\mathbf D}^{(3)}&=&{(y-x)^{2}}\ \Phi_{(2)},\\
{\mathbf D}^{(4)}&=&\sqrt{xy}\ (y-x)\ \Phi_{(2)},\\
{\mathbf D}^{(5)}&=& xy\ \Phi_{(2)}.\\
\end{eqnarray*}
%\end{widetext}
Scattering diagrams are obtained by expanding the Eq. (15) as an infinite series 
and the terms {\bf B}, {\bf F} and {\bf D}$^{(1)}$ to {\bf D}$^{(5)}$ are represented
as scattering vertices (see Fig. 1).
\begin{figure}[h]
\centering
\includegraphics[width=9cm,height=8.5cm]{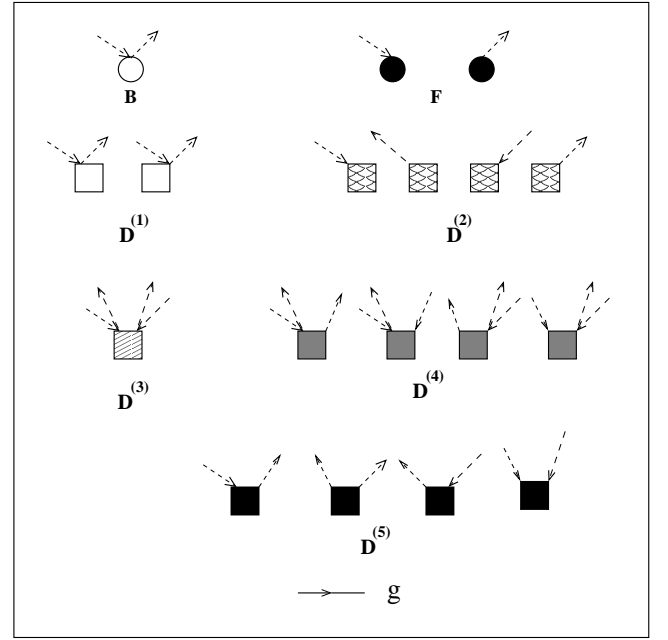}
\caption{The scattering vertices for the averaged Green function }
\label{fig1}
\end{figure}

It might be instructive to understand what these scattering vertices represent physically.
If we look at Eq. (15), we note that the term $\widetilde{\mathbf D}_1$ leads to the
creation or annihilation of {\sl configuration fluctuations} over and above the virtual
crystal description. The vertices {\bf F} shown in Fig 1, create and annihilate a configuration
fluctuation at a given site because of mass disorder, while the vertex {\bf B} counts the
number of such fluctuations at a given site. These are the only type of configuration fluctuations
we can have if we had single site mass disorder alone. The single-site mean-field approximations
like the single-site coherent potential approximation (1CPA) can ideally deal with situations where
we ignore the other vertices in Fig 1.  

The vertices {\bf D}$^{(2)}$ also describe creation and annihilation 
of configuration fluctuations at single sites. That is, fluctuations at any one end of the two-site
dynamical matrix. Similarly, the vertices {\bf D}$^{(1)}$ count the number of configuration fluctuations at any one of the two ends of the dynamical matrix. These are also single-site configuration fluctuations but arise due to fluctuations in the two-site dynamical matrix. 
These may also be treated with some variant of the 1CPA. For example, there are versions of the 1CPA which assume 2{\bf D}$^{AB}$ = {\bf D}$^{AA}$ + {\bf D}$^{BB}$. With such an assumption only the single site configurations fluctuation vertices are non-zero.

The vertices {\bf D}$^{(5)}$ describe creation and annihilation of {\sl two} configuration fluctuations, one at either end of the two-site dynamical matrix. The vertex {\bf D}$^{(3)}$ counts the number
of configuration fluctuations at either end of the dynamical matrix. The vertices {\bf D}$^{(4)}$
are mixed types which both create or annihilate a configuration fluctuation at one end of the dynamical matrix {\sl and} count the number of fluctuations at the other end.

These last three vertices describe configurations fluctuations which are essentially two-site
and cannot be properly described within a single-site mean-field approximation.

\subsection{Scattering diagrams for the averaged Green functions and mean-field
approximations}

\begin{figure}[h]
\centering
\framebox{
\includegraphics[height=7.5cm,width=7.5cm]{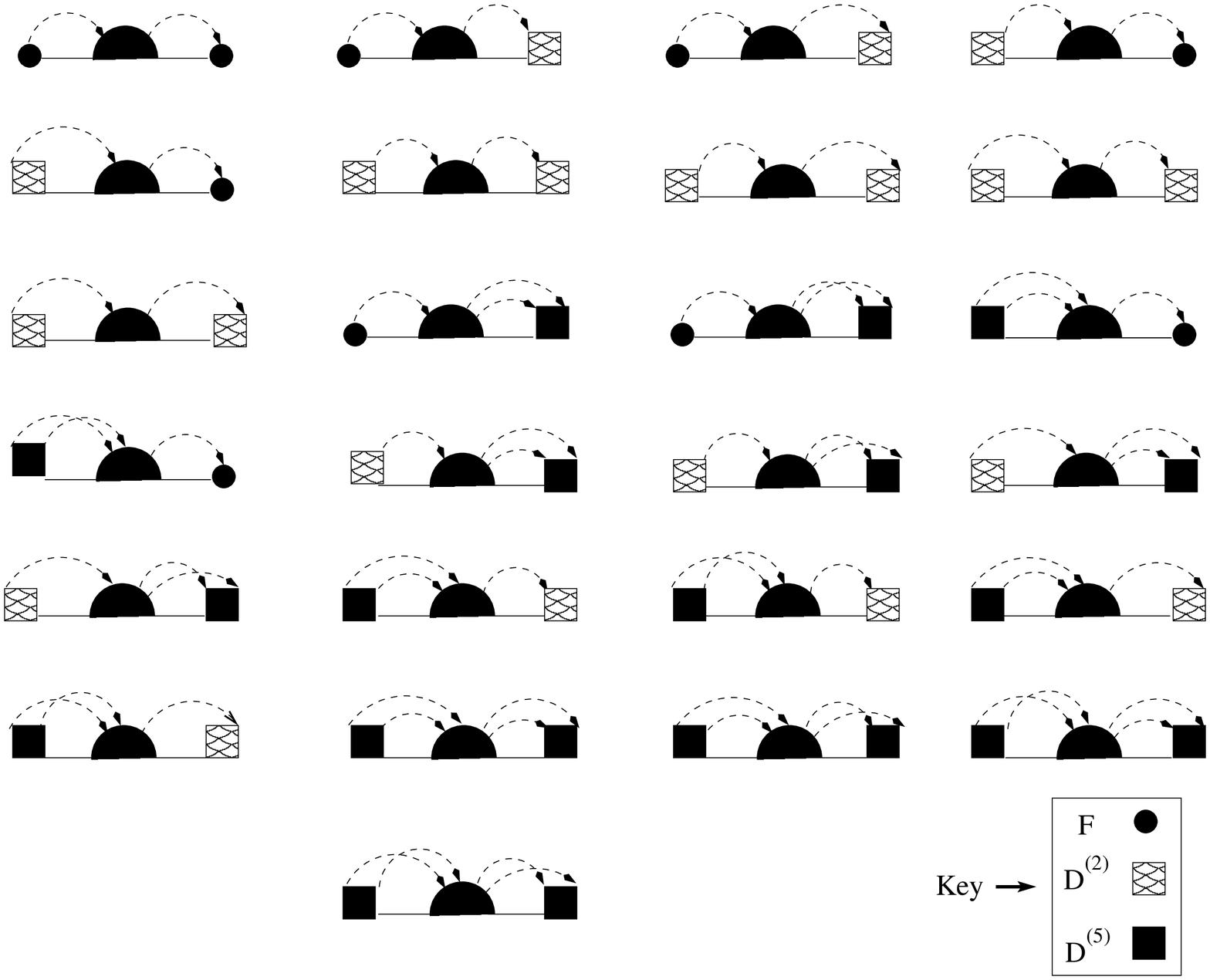}
}
\caption{Structure of the skeleton diagrams for the self-energy}
\label{fig2}
\end{figure}

In our earlier paper \cite{am2} we had developed a 
 multiple scattering picture for the configuration averaged Green function. The idea is very similar to that of Edwards and Langer \cite{ed,la} in the context of purely diagonal disorder. 
The scattering vertices associated with the terms in the 
"perturbing" part of the dynamical matrix : $\wt{\mbf D}_1$  are shown in Fig. \ref{fig1}.

The end point of that formulation was the derivation of a Dyson equation :

\[
\ll {\mbf G}\gg = \g +    \g\ {\mbf \Sigma} \ll {\mbf G}\gg. 
\]
For homogeneous disorder we have shown earlier that we have translational symmetry in the full augmented
space \cite{gdma}. We can then take Fourier transform of the above equation to get

\[  \ll\mbf{G}(\k,E)\gg = \mbf{g}(\k,E) + \mbf{g}(\k,E)\ \mbf{\Sigma}(\k,E)\ \ll\mbf{G}(\k,E)\gg.
\label{dys1}\]

The diagrams for the self-energy are skeleton diagrams\footnote{A skeleton
diagarm is a non-separable diagram all of whose propagators are fully disorder renormalized propagators.}  which have the structure as shown in Fig. \ref{fig2}.
Each of the 25 different  diagram starts with any one of {\bf F}, {\bf D}$^{(2)}$ or {\bf D}$^{(5)}$
and the central dark semicircle represents {\sl all} possible arrangements of scattering vertices to {\sl all}
orders. 

\begin{figure}[b]
\centering
\framebox{
\includegraphics[height=5.5cm,width=7.5cm]{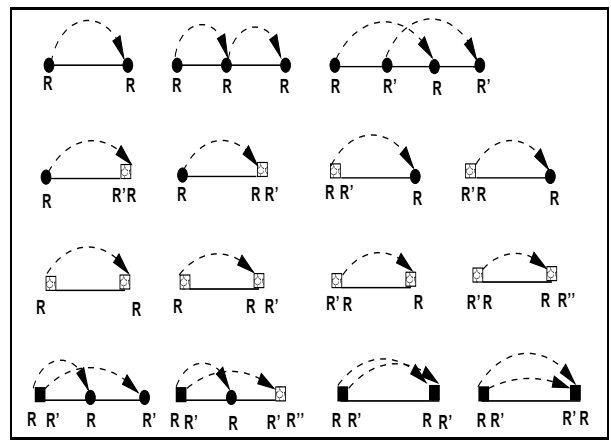}
}
\caption{Details of some skeleton diagrams for the self-energy}
\label{fig3}
\end{figure}

Let us now examine some specific Edwards-Langer scattering diagrams, in some detail, in order to understand
their physical significance and relation to mean-field approximations. The first three diagrams
on the first row of Fig 3 arise because of purely diagonal disorder in mass. Of these the first
two diagrams describe self-energy corrections to the VCA propagator because of configurations
fluctuations at a single site.
These diagrams are closely related to the diagrammatic treatment discussed by
Leath \cite{leath-diag}. Referring to that earlier work, we note that these
diagrams are explicitly included  in the 1CPA.
 The self-energy arising out of such diagrams is site-diagonal, or k-independent in reciprocal space. 
The third diagram is the smallest order diagram of this type which describes joint configuration fluctuations of two distinct sites. These type of diagrams take us beyond the
1CPA. For the 1CPA we ignore these {\sl fourth order} diagrams, and all higher order diagrams, to all orders, 
describing joint configuration fluctuations of more than two sites. 
Within this approximation, the first inaccurate moment 
of the density of states is of order eight. 
If we include diagrams to all orders which describe joint configuration fluctuations of
two sites we are lead  to the
2CPA. This has been described in detail in the work of Aiyer {\sl et.al.} \cite{aekl}
and Nickel and Krummhansl \cite{nk}. 

The diagrams in the second row of Fig 3 describe self-energy corrections due to configuration
fluctuations at a single site for both diagonal (mass) and off-diagonal (force-constant)
disorders. The first and the third diagrams lead to a diagonal self-energy, while the second
and fourth diagrams lead to an off-diagonal contribution to the self-energy in real space. For
off-diagonal disorder even these second order diagrams can lead to off-diagonal (real space) or
k-dependent (reciprocal space) self-energy. Ignoring these contributions
( to all orders)  will lead to a 1CPA
type approximation where even the fourth moment of the density of states will be inaccurate. This had been noted earlier 
 for 1CPA in off-diagonal disorder problems. The diagrams in row three of Fig 3 are very similar contributions from configuration fluctuations at a single site but arising out of
pure off-diagonal disorder.

The diagrams on the last row of Fig 3 describe self-energy contributions coming from joint fluctuations of two sites arising out of off-diagonal disorder. 
All such  diagrams  take us beyond the 1CPA and some of them
contribute to a self-energy which is off-diagonal (real space) or k-dependent (reciprocal space). 

The formal summing up, to infinite order, of diagrams
which involve configuration fluctuations involving single sites has been discussed by Leath \cite{leath-diag}. This leads to the 1CPA and within this approximation the first inaccuracy occurs in the eighth moment of the averaged density of states. Nickel and Krummhansl \cite{nk} have discussed in detail how to sum up, to infinite order, diagrams which also include joint configuration fluctuations of two sites. It is clear from this paper that any generalization of 
a purely diagrammatic treatment of joint multi-site configuration fluctuations is a very difficult task indeed.
However, a recursion based calculation of the self-energy, as proposed by us in our
earlier paper \cite{am2}, includes the contribution of such diagrams. We note that if we calculate
the continued fraction coefficients exactly up to n steps then all the first 2n moments of the density of states are exact. In addition the Beer-Pettifor 
terminator ensures both the herglotz analytic properties of the Green
function and accuracy of the asymptotic moments.  As in our earlier papers
\cite{am1,am2}, therefore, we propose the augmented space recursion as an alternative method for the calculation of averaged Green functions including effects
of joint multi-site configuration fluctuations.

\subsection{Generalized scattering diagrams for the averaged thermal conductivity }

We now go back to equation (\ref{ref1}) and discuss the configuration averaging of the two-particle
Green functions of the kind $\kappa(z_1,z_2)$. The augmented space theorem immediately implies  that 

\begin{eqnarray}
\ll {\mbf \kappa}(z_1,z_2)\gg\phantom{xxxxxxxxxxxxxxxxxxxxxxxxxxxxxx}\nonumber\\
\phantom{x} = \int\frac{d^3\k}{8\pi^3}\ \left\langle  \{\emptyset\} \left |\ \mbox{Tr}\left[ \rule{0mm}{4mm} \right.\wt{\mbf S} \wt{\mbf G}(\k,z_1)\ \wt{\mbf S} \wt{\mbf G}(\k,z_2)\right]\ \left.\rule{0mm}{4mm} \right | \{\emptyset\}\right\rangle\nonumber 
\end{eqnarray}

In real space the random expression $\widehat{\mbf S}$ for a binary alloy can take the values $\widehat{\mbf S}^{AA}$, $\widehat{\mbf S}^{AB}$, $\widehat{\mbf S}^{BA}$ or $\widehat{\mbf S}^{BB}$ with probabilities $x^2$ , $xy$, $yx$ and $y^2$ respectively. We may rewrite the current $\widehat{\mbf S}_{R\alpha,R'\beta}$ as

\begin{eqnarray*}
\widehat{\mbf S}_{R\alpha,R'\beta}= \widehat{\mbf S}^{AA}_{R\alpha,R'\beta}\ n_{R}\ n_{R'} 
 + \widehat{\mbf S}^{AB}_{R \alpha,R'\beta}\ n_{R}\ (1-n_{R'})+ \\
  \pls \widehat{\mbf S}^{BA}_{R \alpha,R'\beta}\ (1-n_{R})\ n_{R'}
+ \widehat{\mbf S}^{BB}_{R \alpha,R'\beta}\ (1-n_{R})(1-n_{R'}).
\end{eqnarray*}

Following the same procedure as for the single particle Green functions we get 

\begin{eqnarray}
 \wt{\S} &=& \sum_{R\alpha}\sum_{R'\alpha'}\left[\rule{0mm}{4mm} \ll \widehat{\S}\gg_{R\alpha,R'\alpha'}\wt{I}\otimes T_{RR'}+{\S}^{(1)}_{R\alpha,R'\alpha'}\right.\nonumber\\
&&\left. \left({ p}^\downarrow_R + { p}^\downarrow_{R'}\right)\otimes T_{RR'}+{\S}^{(2)}_{R\alpha,R'\alpha'} \left({\cal T}^{\uparrow\downarrow}_R + {\cal T}^{\uparrow\downarrow}_{R'}\right)\otimes T_{RR'}\right.\nonumber\\
 &&\left. + {\S}^{(3)}_{R\alpha,R'\alpha'} p^\downarrow_R\otimes p^\downarrow_{R'}\otimes T_{RR'}+ {\S}^{(4)}_{R\alpha,R'\alpha'}\left(p^\downarrow_R\otimes {\cal T}^{\uparrow\downarrow}_{R'}+\right.\right.\nonumber\\ 
&&\left.\left. p^\downarrow_{R'} \otimes{\cal T}^{\uparrow\downarrow}_{R}\right)\otimes T_{RR'}+{\S}^{(5)}_{R\alpha,R'\alpha'} {\cal T}^{\uparrow\downarrow}_R\otimes{\cal T}^{\uparrow\downarrow}_{R'}\otimes T_{RR'}\right],\nonumber\\
\label{eq12}
\end{eqnarray}

\n where

\begin{eqnarray*}
\S^{(1)}&=&(y-x)\ \widehat{\S}^{(1)},\ \  \S^{(2)} = \sqrt{xy}\ \widehat{\S}^{(1)}, \\ \S^{(3)} &=& (y-x)^2\ \widehat{\S}^{(2)},\ \ \S^{(4)} =\sqrt{xy}\ (y-x)\ \widehat{\S}^{(2)},\\ \S^{(5)} &=& xy\ \widehat{\S}^{(2)}.
\end{eqnarray*}

\n and

\begin{eqnarray*}
 \widehat{\S}^{(1)}_{R\alpha,R'\beta} &=& x \left( \widehat{\S}^{AA}_{R\alpha,R'\beta}- \widehat{\S}^{AB}_{R\alpha,R'\beta}\right) -  y \left( \widehat{\S}^{BB}_{R\alpha,R'\beta}-\right. \\ && \left. \widehat{\S}^{BA}_{R\alpha,R'\beta}\right)\\
 \widehat{\S}^{(2)}_{R\alpha,R'\beta}&=&\widehat{\S}^{AA}_{R\alpha,R'\beta} + \widehat{\S}^{BB}_{R\alpha,R'\beta} - \widehat{\S}^{AB}_{R\alpha,R'\beta} - \widehat{\S}^{BA}_{R\alpha,R'\beta}. 
\end{eqnarray*}

\begin{figure}[h]
\centering
\includegraphics[height=8cm,width=8cm]{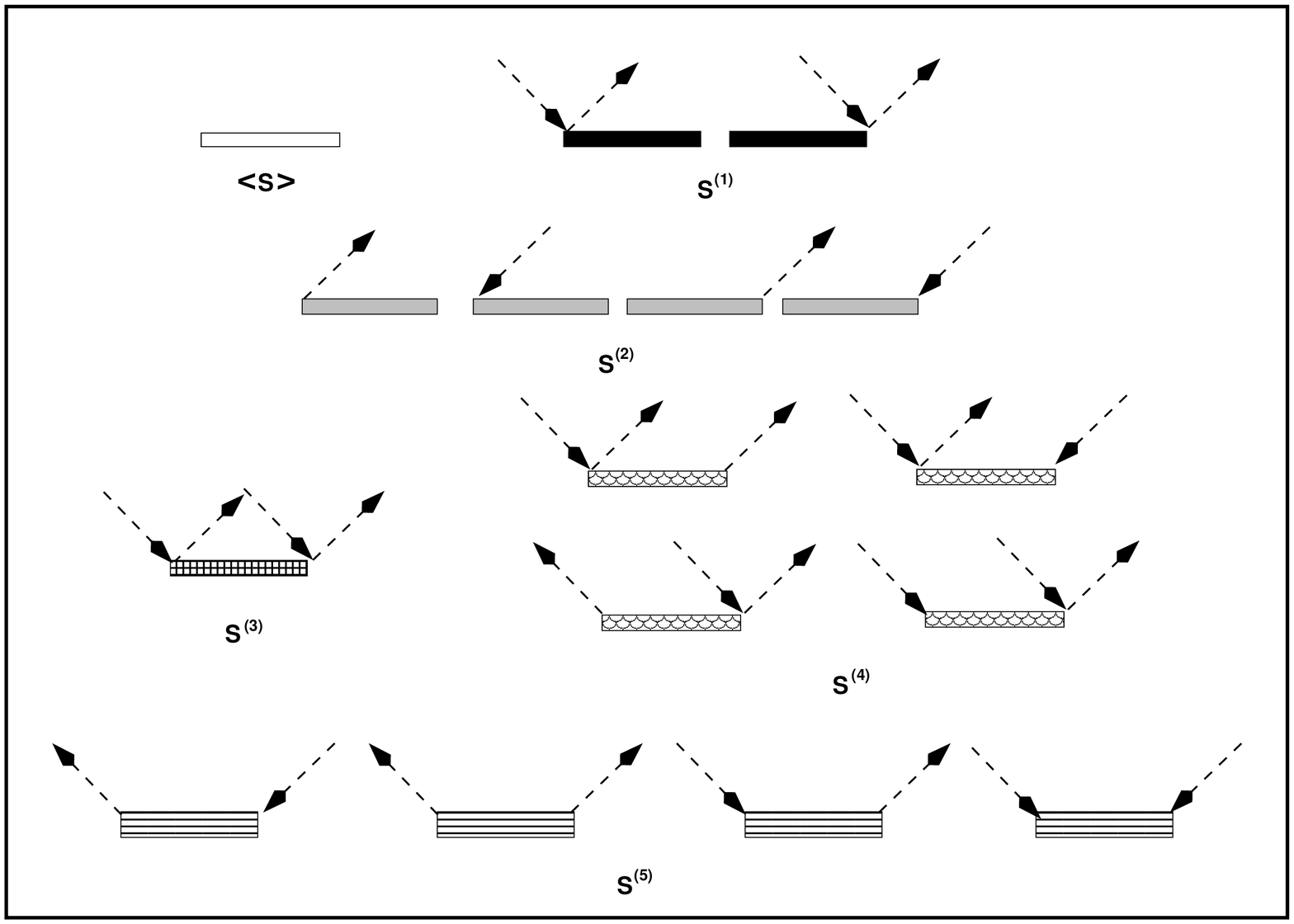}
\caption{The scattering vertices associated with the random current terms}
\label{fig4}
\end{figure}

Equation (\ref{eq12}) is very similar to (\ref{dyn}), which shows the terms arising in augmented space due to the disorder in the dynamical matrix. Fig. \ref{fig4} shows the sixteen different scattering vertices arising out of Eq. (\ref{eq12}). We may compare the scattering vertices arising out of disorder in the current term with those that arise from disorder in the dynamical matrix shown in Fig. 1. We
note the close similarity between them leading to both terms which arise because of configuration fluctuations at a single site and also two sites.

Note that the averaging of $\kappa(z_1,z_2)$ involves averaging of a product of four terms
two involving the thermal currents $S$ and two Green functions. These random functions are
not independent, but are all intrinsic functions of the random occupation variables $n_R$. The
average of the product is therefore {\sl not} the product of the averages. 

\begin{figure}[h]
\centering
\includegraphics[height=6.4cm,width=8cm]{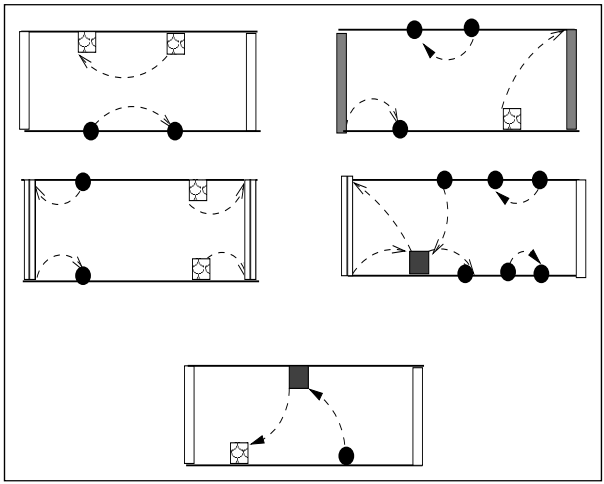}
\caption{Different classes of scattering diagrams for the thermal conductivity }
\label{fig5}
\end{figure}

Let us now examine the various classes of scattering diagrams. The simplest one is the type shown
on the top left of Fig \ref{fig5}. These diagrams involve two averaged current terms
and two propagators decorated with scattering diagrams in all possible ways. The contribution of such a term (the zero-th order term) is then,

\begin{eqnarray*}
\kappa_{(0)}(z_1,z_2) &=& \int \frac{\displaystyle d^3\k}{\displaystyle 8\pi^3}\  \mathrm{Tr} \left[\rule{0mm}{4mm}\ll {\mathbf S}\gg
\ll {\mathbf G}(\k,z_1)\gg\right.\nonumber\\
& & \left.\rule{0mm}{4mm}  \ll {\mathbf S}\gg {\mathbf G}(\k,z_2)\gg \right]
\end{eqnarray*}

\begin{figure}[h]
\centering
\includegraphics[height=2in,width=3in]{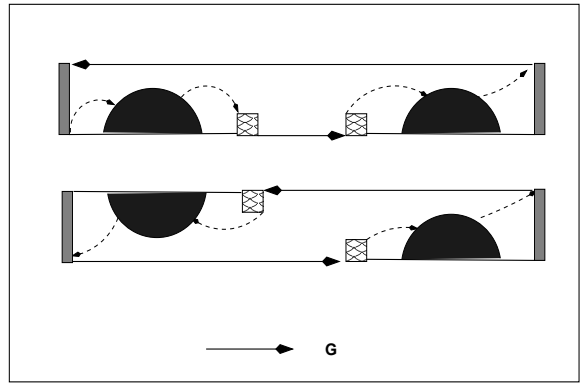}
\caption{Two examples of scattering diagrams where no disorder line joins the two phonon propagators}
\label{fig6}
\end{figure}

\begin{figure*}[t]
\centering
\includegraphics[height=3.0in,width=3.5in]{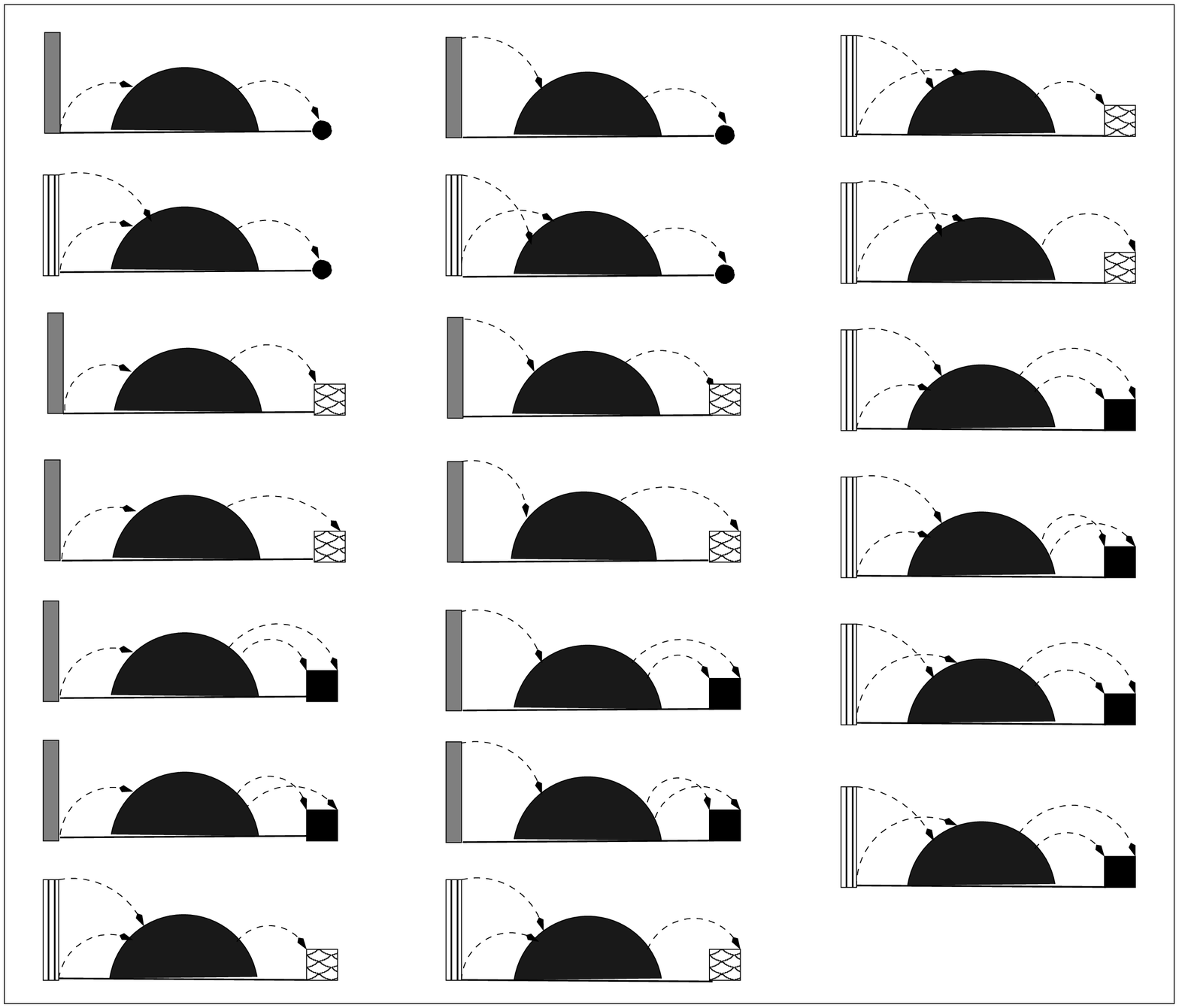}
\includegraphics[height=3.0in,width=3.5in]{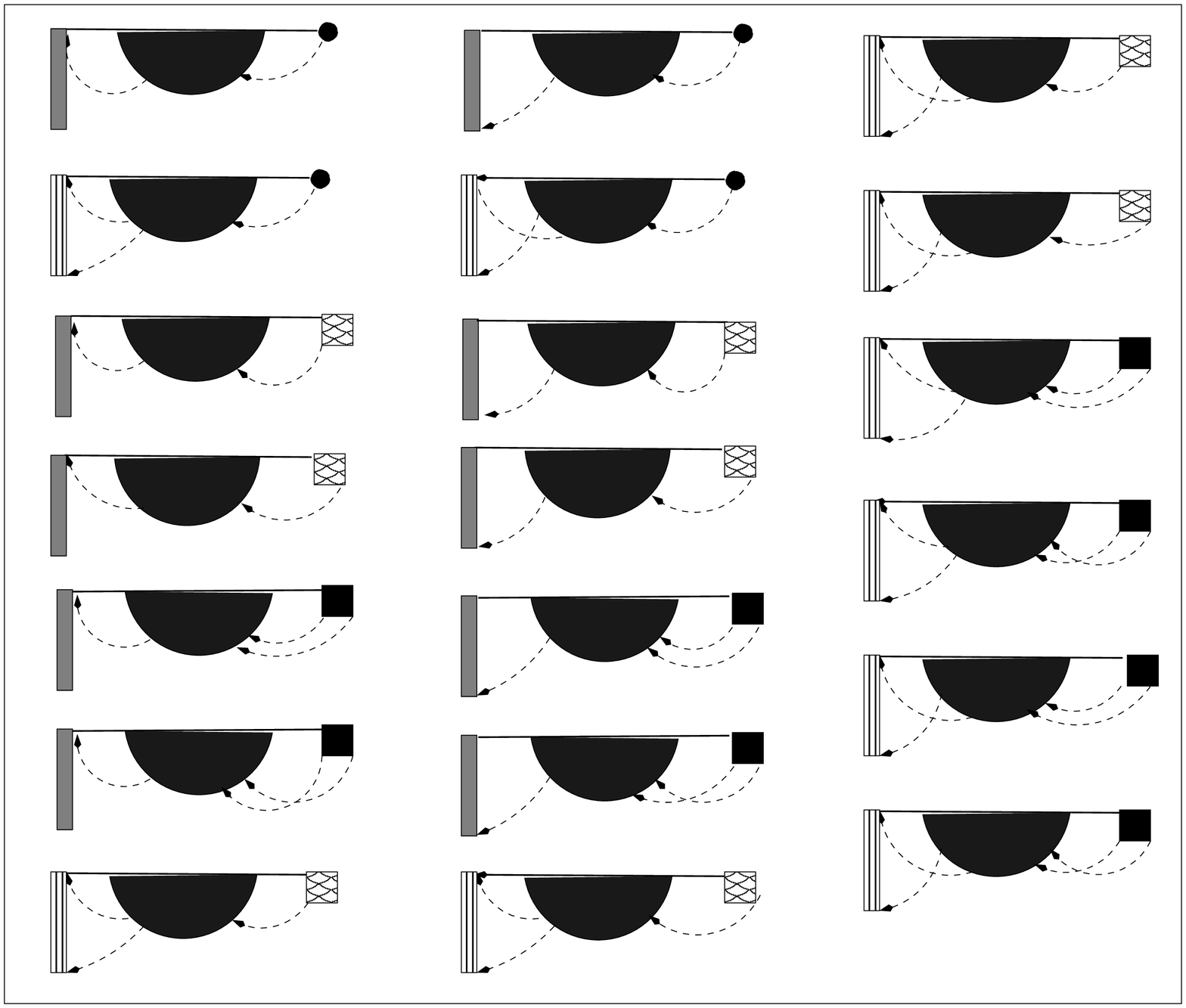}
\caption{Scattering diagrams contributing to effective heat current}
\label{fig7}
\end{figure*}

The next type of scattering diagrams are the types shown on the top right of Fig \ref{fig5}.
These diagrams connect a current term to a propagator. 
 The first type of scattering diagrams are those in which disorder lines do not connect the two phonon propagators. Fig \ref{fig6} shows the general structure of such diagrams. It is
clear from the figure that these diagrams renormalize the current terms. Fig \ref{fig7}
shows all possible renormalized currents arising out of diagrams of this class.

 In our earlier paper on optical conductivity \cite{sm} we had shown that this type of diagrams provide the predominant disorder correction to the current terms in the electronic
problem. If we compare them with the diagrams for the self energy (Fig. \ref{fig2}) we note that for the diagrams in the left panel of Fig. \ref{fig7}, the only difference is that the leftmost scattering vertex is replaced by a very similar current term. Similarly the diagrams on the right panel are the same as the self-energy diagrams except that the last vertex is replaced by a very similar current term. In all these diagrams of Fig. \ref{fig7} the left and rightmost diagonal term similar to the 
 vertex {\bf F} of Fig. \ref{fig1} is of course missing. We may then go on to write 

\begin{eqnarray}
\Delta {\S}_1(\k)(z_1,z_2) =\phantom{xxxxxxxxxxXXXXXXXX} \nonumber\\
 2 \left(\rule{0mm}{3mm} {\S}^{(2)}(\k) + {\S}^{(5)}(\k)\right)\left[\rule{0mm}{3.5mm}\Delta{(\k,z_1)}\right]^{-1} {\mbf \Sigma}(\k,z_1)+\nonumber\\
{\mbf \Sigma}(\k,z_2)\left[\rule{0mm}{3.5mm}\Delta{(\k,z_2)}\right]^{-1} 2\left(\rule{0mm}{3mm}  {\S}^{(2)}(\k) + {\S}^{(5)}(\k)\right),\phantom{x}\nonumber\\
\end{eqnarray}

\begin{figure}[b]
\centering 
\includegraphics[height=0.7in,width=2.7in]{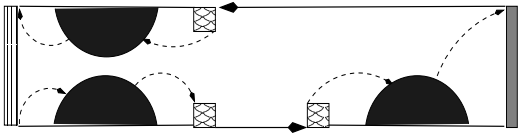}
\caption{More scattering diagrams contributing to effective current}
\label{fig8}
\end{figure}
where
\[
\Delta{(\k,z)}={\mbf F}z^2 + 2 {\mbf D}^{(2)}(\k) +2 {\mbf D}^{(5)}(\k). 
\]

The next most dominant disorder corrections come from the diagrams shown in the middle row
of Fig \ref{fig5}. A general such diagram is  shown in Fig. \ref{fig8}. These diagram
connect one current term to two propagators. These diagrams also renormalize this
current term. 
Again looking at these diagrams we note that they are also related to a self-energy diagrams
with vertices at both ends replaced by currents. We therefore may write
\begin{eqnarray}
\Delta {\S}_2(\k)(z_1,z_2) = \phantom{XXXXXXXXXXXXXXXXX}\nonumber\\
 {\mbf \Sigma}(\k,z_2)\left[\rule{0mm}{3.5mm}\Delta{(\k,z_2)}\right]^{-1}{\mbf S}^{(5)}(\k) \left[\rule{0mm}{3.5mm}\Delta{(\k,z_1)}\right]^{-1}{\mbf  \Sigma}(\k,z_1)\nonumber\\
 + {\mbf \Sigma}(\k,z_1)\left[\rule{0mm}{3.5mm}\Delta{(\k,z_1)}\right]^{-1}{\mbf S}^{(5)}(\k) \left[\rule{0mm}{3.5mm}\Delta{(\k,z_2)}\right]^{-1}{\mbf  \Sigma}(\k,z_2).\nonumber\\
\end{eqnarray}

\n The effective current is then given by 

\begin{eqnarray}
{\S}_{\mathrm{eff}}(\k,z_1,z_2)  =\phantom{XXXXXXXXXXXXXX} \nonumber\\
\ll \widehat{\S}(\k)\gg + \Delta{\S}_1(\k)(z_1,z_2) + \Delta{\S}_2(\k)(z_1,z_2).
\end{eqnarray}

In our earlier paper \cite{sm} on a similar electronic problem we have argued that these are the dominant disorder corrections to the
current term. With these corrections we obtain 
\begin{eqnarray}
\kappa_{(1)}(z_{1},z_{2})&=& \int\frac{d^3\k}{8\pi^3}\ \mathrm{Tr}\left[\rule{0mm}{4mm}{\S}_{\mathrm{eff}}(\k,z_{1},z_{2})\ll {\mbf{G}}({\bf k},z_{1})\gg\right.\nonumber\\
&&\left. {\S}_{\mathrm{eff}}^{\dagger}(\k,z_{1},z_{2})\ll {\mbf{G}}({\bf k},z_{2})\gg\right].
\label{xyz}
\end{eqnarray}

Since averaging the thermal conductivity involves averaging the product of two
propagators and two current terms, we may physically understand the different terms in Fig. 5 in the following way : the top left diagram is related to the
product of the averages. The term on the top row to the right is related to
a pair of  joint averages of one propagator and one current term. The diagrams
on the second row are related to the product of the average of one current and
the average of the product of two propagators and a current. The bottom most
diagram is related to the product of two averaged currents and the average
of the product of two propagators.

\subsection{Averaged thermal diffusivity}
For a harmonic solid, a temperature independent mode diffusivity $D_{\gamma}$ is defined as 
\[
D^{\mu\nu}_{\gamma}(\k)=\pi \sum_{\gamma^{\prime}\ne\gamma}\frac{1}{\omega_{\k\gamma}^{2}}{\mbf S}^{\mu}_{\gamma\gamma^{\prime}}(\k){\mbf S}^{\nu}_{\gamma^{\prime}\gamma}(\k)\ \delta(\omega_{\k\gamma}-\omega_{\k\gamma^{\prime}}).
\]
This is an intrinsic property of the $\gamma$-th normal mode and provides an unambiguous criterion for localization.

The averaged thermal diffusivity (averaged over modes) is then given by
\begin{eqnarray}
{\bf D}^{\mu\nu}(\omega)&=&\frac{\displaystyle\int\frac{d^{3}\k}{8 \pi^{3}}\sum_{\gamma}D^{\mu\nu}_{\gamma} (\k)\delta(\omega - \omega_{\k\gamma})}{\displaystyle\int\frac{ d^{3}\k}{8 \pi^{3}}\sum_{\gamma}\delta(\omega - \omega_{\k\gamma})}\nonumber\\
&=& \frac{D^{\mu\nu}_{tot}(\omega)}{{\displaystyle\int\frac{ d^{3}\k}{8 \pi^{3}}\sum_{\gamma}\delta(\omega - \omega_{\k\gamma})}}.
\end{eqnarray}
 Assuming isotropy of the response, we can rewrite the numerator of above equation as

\begin{eqnarray}
D^{\mu\mu}_{tot}(\omega)&=&\pi\int d\omega^{\prime}\int \frac{d^{3}\k}{8 \pi^{3}}\sum_{\gamma}\sum_{\gamma^{\prime}}\widehat{S}^{\mu}_{\gamma\gamma^{\prime}}(\k)\widehat{S}^{\mu}_{\gamma^{\prime}\gamma}(\k)\nonumber\\
&& \delta(\omega^{\prime}-\omega_{{\k}\gamma^{\prime}})\delta(\omega_{{\k}\gamma}-\omega^{\prime})\delta(\omega-\omega_{{\k}\gamma}),\nonumber
\end{eqnarray}
\noindent where
\[
\widehat{S}^{\mu}_{\gamma\gamma^{\prime}}(\k)\ =\ \frac{1}{\omega_{\k \gamma}}{\mbf S}^{\mu}_{\gamma\gamma^{\prime}}(\k).
\]
We may again rewrite the above equation for $D_{tot}$ as
\begin{eqnarray}
D^{\mu\mu}_{tot}(\omega)&=&\frac{1}{\pi^{2}}\int d\omega^{\prime}\int \frac{d^{3}\k}{8 \pi^{3}} \mbox{Tr} \left[ \Im m \{{\mbf G}(\k,\omega^{\prime})\} \widehat{\mbf S}^{\mu}(\k) \right.\nonumber\\
&& \left. \hspace{-0.4cm} \Im m \{{\mbf G}(\k,\omega^{\prime})\}\widehat{\mbf S}^{\mu}(\k)\Im m \{{\mbf G}(\k,\omega)\}  \right].\nonumber
\end{eqnarray}
 The averaged thermal diffusivity can then be expressed as (\ for an isotropic response\ )
\begin{eqnarray}
{\mbf D}(\omega)&=&\frac{1}{3}\sum_{\mu}D^{\mu\mu}(\omega)\nonumber\\
&=&\frac{\pi}{3}\frac{\sum_{\mu}{\mbf D}_{tot}^{\mu\mu}(\omega)}{\displaystyle\int \frac{d^{3}\k}{8\pi^{3}}\mbox{Tr}\left[\Im m \{{\mbf G}(\k,\omega)\}\right]}.
\end{eqnarray}
For disordered material, we shall be interested as before in obtaining the configuration averaged thermal diffusivity. We have already discussed the configuration averaging of the two particle Green function using scattering diagram technique in Sec. III(C). It has been found that the net effect is to replace the current terms by an effective heat current ${\mbf S}_{eff}(\k,z_1,z_2)$. The effective current is a sum of average current and the terms arising out of the disorder correction.  As in the case of optical conductivity calculation \cite{sm}, it will be shown that the overall contribution of disorder correction terms to the thermal conductivity is very small as compared to the average current $\ll\widehat{\mbf S}(\k)\gg$. Keeping in mind this effect of disorder correction terms to the heat current, the configuration averaged thermal diffusivity can be expressed (to a 1st order approximation) in the form
\begin{equation}
\ll{\mbf D}(\omega)\gg =\frac{\pi}{3}\frac{\sum_{\mu}\ll{\mbf D}_{tot}^{\mu\mu}(\omega)\gg}{\displaystyle\int \frac{d^{3}\k}{8\pi^{3}}\mbox{Tr}\left[\Im m \ll{\mbf G}(\k,\omega)\gg\right]},
\end{equation}
\noindent where
\begin{eqnarray}
\ll D^{\mu\mu}_{tot}(\omega)\gg\ \simeq\ \frac{1}{\pi^{2}}\hspace{-0.1cm}\int\hspace{-0.2cm} d\omega^{\prime}\hspace{-0.1cm}\int \frac{d^{3}\k}{8 \pi^{3}} \mbox{Tr} \left[\rule{0cm}{4mm} \Im m \ll{\mbf G}(\k,\omega^{\prime})\gg \right.\nonumber\\ 
\left.\ll \widehat{\mbf S}^{\mu}(\k)\ \Im m\ \{{\mbf G}(\k,\omega^{\prime})\} \ \widehat{\mbf S}^{\mu}(\k)\ \rule{0cm}{4mm} \Im m \{{\mbf G}(\k,\omega)\}\gg  \right].\nonumber
\end{eqnarray}
\begin{equation}
\end{equation}
\subsection{The vertex correction}
\begin{figure}[h]
\centering
\includegraphics[width=8cm,height=2cm]{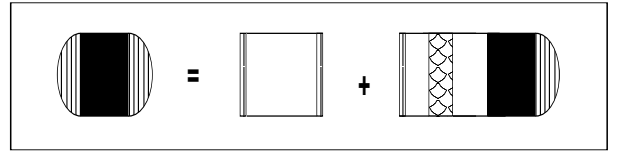}
\caption{The Bethe-Salpeter equation for the thermal conductivity}
\label{fig9}
\end{figure}

We shall now examine the scattering diagrams we have left out, namely, those in which disorder lines connect both the propagators directly. One such diagram is the bottom-most diagram shown
in Fig \ref{fig5}. These lead to vertex corrections due to correlated propagation. 
In general we obtain a Bethe-Salpeter equation for the averaged two-particle propagator.
This is diagrammatically shown in Fig \ref{fig9}. 

To obtain the vertex correction, shown in Fig \ref{fig9}, we 
shall consider only a special class of vertex diagrams in this communication : namely the scattering diagrams which are built out of repeated vertices. These are called the ladder diagrams and can be summed up to all orders. This is the disorder scattering version of the random phase approximation (RPA).

\begin{figure}[t]
\centering
\includegraphics[width=3.5in,height=3.2in]{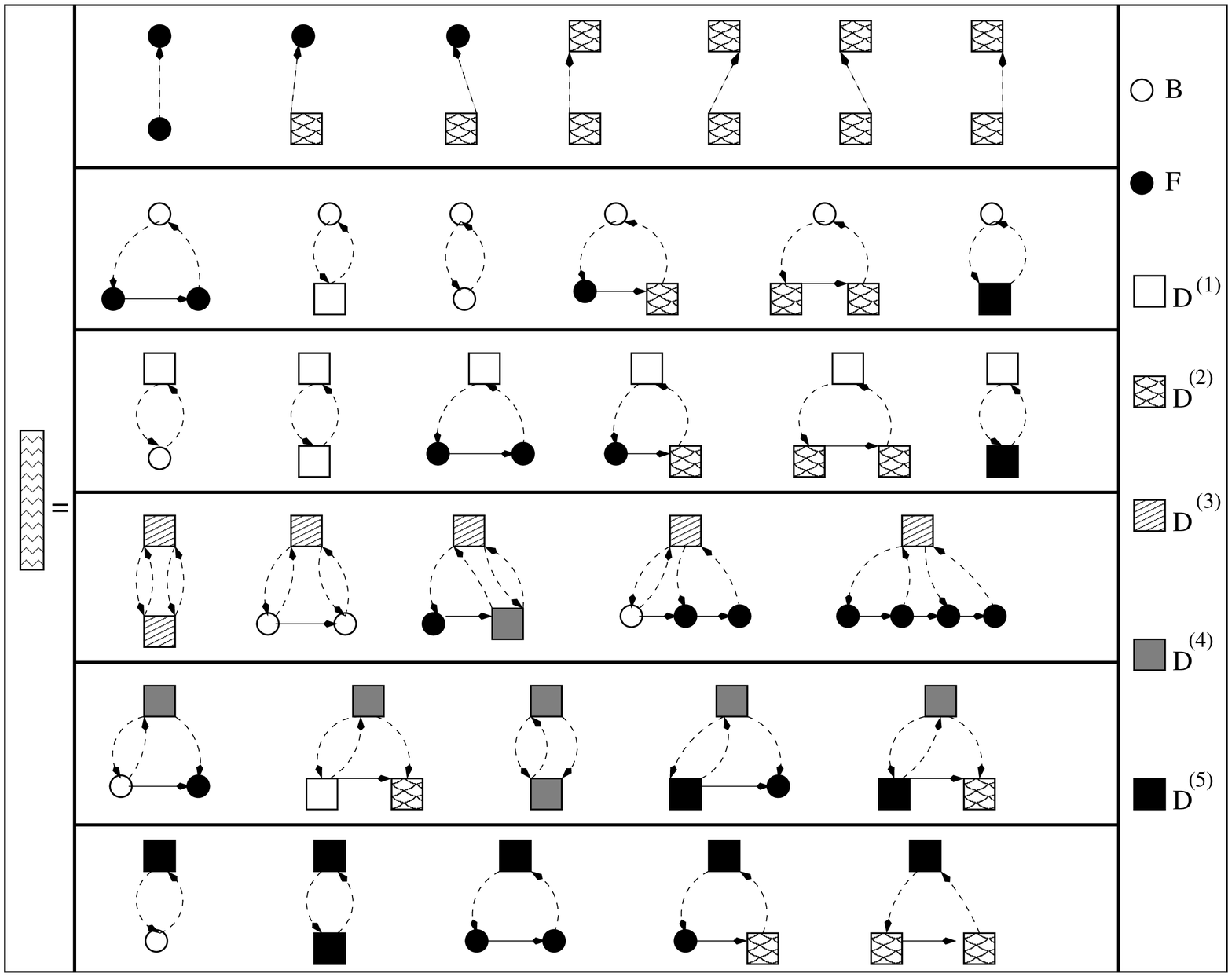}
\caption{The scattering diagrams for the ladder type of vertex correction.} 
\label{fig10}
\end{figure}

A few amongst all the possible scattering diagrams for the ladder type of vertex corrections involving the vertices ${\mbf B}$, ${\mbf F}$, ${\mbf D}^{(1)}$, ${\mbf D}^{({2})}$, ${\mbf D}^{({3})}$, ${\mbf D}^{({4})}$, ${\mbf D}^{({5})}$ are as shown in Fig. \ref{fig10}

The contribution of diagrams from the first line of Fig. \ref{fig10} in the four legged vertex (shown in the extreme left column of the same figure) is given by
\begin{eqnarray}
 \left(W^{\gamma\delta}_{\alpha\beta}\right)_{\mbox{1st line}} =\phantom{xxxxxxxxxxxxxxxxxxxxxx}\nonumber\\
 (z_1z_2)^2\ F_\alpha F_\gamma\ \delta_{\alpha\beta}\delta_{\gamma\delta} + 2 z_2^2\ D^{(2)}_{\alpha\beta}\ F^\gamma\ \delta_{\gamma\delta}\nonumber\\  
+ 2 z_1^2\ D^{(2)}_{\gamma\delta}\ F^\alpha\ \delta_{\alpha\beta}  
+ 4 D^{(2)}_{\alpha\beta}\ D^{(2)}_{\gamma\delta}.
\end{eqnarray}
The contribution of the second line of diagrams and all those obtained from them by simple
reflection operations about the vertical and horizontal directions is given by 
\begin{widetext}
\begin{eqnarray}
 \left(W^{\gamma\delta}_{\alpha\beta}\right)_{\mbox{2nd line}} =({z_1}^{2}z_2)^{2} \left[\sum_{\nu^{\prime}\nu^{\prime\prime}}( F_{\alpha\nu^{\prime}} \delta_{\alpha\nu^{\prime}})\ G_{R\nu^{\prime},R\nu{\prime\prime}}\ ( F_{\nu^{\prime\prime}\beta} \delta_{\nu^{\prime\prime}\beta})\right](B_{\nu\delta}\delta_{\nu\delta}) +
({z_1}{z_2}^{2})^{2} (B_{\alpha\beta}\delta_{\alpha\beta})\nonumber\\
\times \left[\sum_{\nu^{\prime}\nu^{\prime\prime}}( F_{\nu\nu^{\prime}} \delta_{\nu\nu^{\prime}})\ G_{R\nu^{\prime},R\nu{\prime\prime}}\ ( F_{\nu^{\prime\prime}\delta} \delta_{\nu^{\prime\prime}\delta})\right]+ 
2\left[D^{(1)}_{\alpha\beta}({z_2}^{2}B_{\nu\delta}\delta_{\nu\delta} )\right] + 2\left[({z_1}^{2}B_{\alpha\beta}\delta_{\alpha\beta} ) D^{(1)}_{\nu\delta}\right] + ({z_1}^{2}B_{\alpha\beta}\delta_{\alpha\beta}) ({z_2}^{2}B_{\nu\delta}\delta_{\nu\delta})
\nonumber\\
+\ 2 \left[\sum_{\nu^{\prime}\nu^{\prime\prime}}({z_1}^{2} F_{\alpha\nu^{\prime}} \delta_{\alpha\nu^{\prime}})\ G_{R\nu^{\prime},R\nu{\prime\prime}}\  D^{(2)}_{\nu^{\prime\prime}\beta} \right]({z_2}^{2}B_{\nu\delta}\delta_{\nu\delta}) +
 2\ ({z_1}^{2}B_{\alpha\beta}\delta_{\alpha\beta}) \left[\sum_{\nu^{\prime}\nu^{\prime\prime}}({z_2}^{2} F_{\nu\nu^{\prime}} \delta_{\nu\nu^{\prime}})\ G_{R\nu^{\prime},R\nu{\prime\prime}}\  D^{(2)}_{\nu^{\prime\prime}\delta} \right]\nonumber\\
 +\ 2 \left[\sum_{\nu^{\prime}\nu^{\prime\prime}} D^{(2)}_{\alpha\nu^{\prime}}\ G_{R\nu^{\prime},R\nu{\prime\prime}}\ ({z_1}^{2} F_{\nu^{\prime\prime}\beta} \delta_{\nu^{\prime\prime}\beta}) \right]({z_2}^{2}B_{\nu\delta}\delta_{\nu\delta}) +
 2\ ({z_1}^{2}B_{\alpha\beta}\delta_{\alpha\beta}) \left[\sum_{\nu^{\prime}\nu^{\prime\prime}} D^{(2)}_{\nu\nu^{\prime}}\ G_{R\nu^{\prime},R\nu{\prime\prime}}\ ({z_2}^{2} F_{\nu^{\prime\prime}\delta} \delta_{\nu^{\prime\prime}\delta}) \right] \nonumber\\
 +\ 4 \left[\sum_{\nu^{\prime}\nu^{\prime\prime}} D^{(2)}_{\alpha\nu^{\prime}}  \ G_{R\nu^{\prime},R\nu{\prime\prime}}\  D^{(2)}_{\nu^{\prime\prime}\beta} \right]({z_2}^{2}B_{\nu\delta}\delta_{\nu\delta}) + 4 ({z_1}^{2}B_{\alpha\beta}\delta_{\alpha\beta})\left[\sum_{\nu^{\prime}\nu^{\prime\prime}} D^{(2)}_{\nu\nu^{\prime}}\ G_{R\nu^{\prime},R\nu{\prime\prime}}\  D^{(2)}_{\nu^{\prime\prime}\delta} \right] \phantom{XXXXXXXX}\nonumber\\
+\ D^{(5)}_{\alpha\beta}({z_2}^{2}B_{\nu\delta}\delta_{\nu\delta}) + ({z_1}^{2}B_{\alpha\beta}\delta_{\alpha\beta})D^{(5)}_{\nu\delta}.\phantom{XXXXXXXXXXXXXXXXXXXXXXXXXXXX}
\end{eqnarray}
\end{widetext}
\n The contribution of rest of the diagrams may be obtained in exactly the same way.

\begin{figure}
\centering
\includegraphics[width=3.5in,height=1cm]{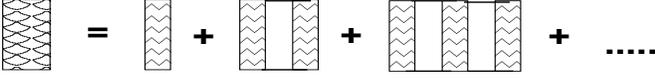}
\caption{The structure of infinite series of ladder diagrams.} 
\label{fig11}
\end{figure}

\n If the sum of all possible scattering diagrams contributing to the four legged vertex shown in the left column of Fig. \ref{fig10} is denoted by $W_{\alpha\beta}^{\nu\delta}$, then the infinite series of ladder diagrams shown in Fig. \ref{fig11} may be directly summed up as follows : If we define

\begin{eqnarray*}
{\mbf \Gamma}(z_1,z_2) & = &\int_{BZ}\frac{d^3\k}{8\pi^3}\ {\mbf G}(\k,z_2)\widehat{\S}^{\mathrm{eff}}(\k,z_1,z_2) {\mbf G}(\k,z_1),\\
\widehat{\mbf \Gamma}(z_1,z_2)&  = &\int_{BZ}\frac{d^3\k}{8\pi^3}\ {\mbf G}(\k,z_1)\widehat{\S}^{\mathrm{eff}}(\k,z_1,z_2)^\dagger\ {\mbf G}(\k,z_2),\\
{\mbf \Theta}_{\alpha\beta}^{\gamma\delta}(z_1,z_2) & = &  \int_{BZ}\frac{d^3\k}{8\pi^3}\ G_{\alpha\beta}(\k,z_1)\ G_{\gamma\delta}(\k,z_2), \\
\end{eqnarray*}

Then

\begin{eqnarray}
{\mbf \Lambda}(z_1,z_2)& = & {\mbf W} + {\mbf W} {\mbf \Theta} {\mbf W} + {\mbf W} {\mbf \Theta} {\mbf W}  {\mbf \Theta} {\mbf W}+ \ldots \nonumber\\
& =& {\mbf W}(z_1,z_2) \left(\rule{0mm}{4mm}{\mbf I}-{\mbf \Theta}(z_1,z_2) {\mbf W}(z_1,z_2)\right)^{-1}. \nonumber\\
\end{eqnarray}

Thus we get 
\begin{eqnarray}
\ll \Delta \kappa(z_1,z_2)^{\mathrm ladder}\gg \phantom{XXXXXXXXXXXXX}\nonumber\\
= \sum_{\alpha\beta}\sum_{\nu\delta}{\mbf \Gamma}^{\alpha}_{\beta}(z_1,z_2)\ {\mbf \Lambda}_{\beta\delta}^{\alpha\nu}(z_1,z_2)\ \widehat{\mbf \Gamma}^{\nu}_{\delta}(z_1,z_2)\phantom{xxx} \nonumber\\
=  \mathrm{Tr} \left[\rule{0mm}{4mm}{\mbf \Gamma}(z_1,z_2)\otimes \widehat{\mbf \Gamma}(z_1,z_2)\ {\mbf \odot}\ {\mbf \Lambda}(z_1,z_2)\right].\phantom{xxxx} 
\end{eqnarray}

The configuration average of the two-particle green function is then given by
\begin{equation}
\ll \kappa(z_1,z_2) \gg = \ll\kappa_{(1)}(z_1,z_2) \gg\ +\ \ll \Delta \kappa(z_1,z_2)^{\mathrm ladder}\gg
\end{equation}

\section{DETAILS OF NUMERICAL IMPLEMENTATION}

Although we have used the scattering diagrammatic technique to analyze
the effects of disorder scattering on the thermal conductivity and obtain
relation between the effective current and self-energy, we shall not use
this approach to actually numerically obtain the thermal conductivity for 
a real alloy system. If we look at the earlier sections we note that what
we need to obtain are essentially 
the configuration averaged Green matrices and the
matrix self-energies. For this we shall use the  augmented space block-recursion
 \cite{am2} and also the Brillouin zone integration scheme developed by us earlier \cite{smj}. In this section we shall briefly discuss
the two techniques.

\subsection{Block Recursion}

As is clear from the expression of configuration averaged lattice thermal conductivity and thermal diffusivity that the numerical evaluation of these quantities require the full Green matrix and the self energy matrix instead of their diagonal elements alone. 

In a recent communication \cite{am2}, we have already described the block recursion technique, which calculates the entire green matrix and the self energy matrix. For the sake of completeness, we shall describe in brief the salient features of the technique. In the block recursion, we start from a matrix basis of the form  $\{\Phi^{(n)}_{J,\alpha\beta}\}$, where $J$ is the discrete labeling of the augmented space states and the $\alpha,\beta$ labels Cartesian directions.

For a reciprocal space based calculation (as in the present case) we start with 

\[ 
\Phi^{(1)}_{J,\alpha\beta}\ =\ U_{\alpha\beta}^{(1)} \ \delta_{J,1} + U_{\alpha\beta}^{(2)} \ \delta_{J,2}, 
\]

\n where

$U_{\alpha\beta}^{(1)}=\frac{\displaystyle A(m^{-1/2})}{\displaystyle\left[\rule{0mm}{2mm}A(m^{-1})\right]^{1/2}}\ \delta_{\alpha\beta}, \ \ \ U_{\alpha\beta}^{(2)}=\frac{\displaystyle F(m^{-1/2})}{\displaystyle\left[\rule{0mm}{2mm}A(m^{-1})\right]^{1/2}}\ \delta_{\alpha\beta}$,

\n with

$A(X)=(x X_{A}+y X_{B}),\ \ \ F(X) = (y-x)(X_A-X_B)$. 
\vskip 0.3cm

The remaining terms in the basis are recursively obtained from 

\begin{eqnarray}
\sum_{\beta '} \Phi^{(n+1)}_{J,\alpha\beta '} B^{(n+1)\dagger}_{\beta '\beta}& = &\sum_{J'}\sum_{\beta'} \wt{H}_{J\alpha,J'\beta '} \Phi^{(n)}_{J',\beta '\beta}- \nonumber\\
&& \sum_{\beta '}\Phi^{(n)}_{J,\alpha\beta '} A^{(n)}_{\beta '\beta} - \sum_{\beta '}\Phi^{(n-1)}_{J,\alpha\beta '} B^{(n)}_{\beta '\beta},\nonumber
\end{eqnarray}

where $\wt H$ is the lattice vibrational Hamiltonian (under harmonic approximation) in the augmented space. We then use the Gram-Schmidt orthonormalization method to keep track of orthogonality condition among the various columns of $\Phi^{(n)}_{J,\alpha\beta}$ and hence to calculate the matrix coefficient ${\bf B}^{(n+1)}$. Since the Hamiltonian in this new basis is block tri-diagonal, so the green matrix can be expressed as a matrix continued fraction in terms of the coefficients ${{\bf A}^{n},{\bf B}^{(n+1)}}$ as follows :
\begin{eqnarray}
{\mbf G}^{(n)}&=& \left[\rule{0mm}{4mm} w^2\ {\mbf I}- {\mbf A}^{(n)}-
{\mbf B}^{(n+1)\dagger}\ {\mbf G}^{(n+1)}\ {\mbf B}^{(n+1)}\right]^{-1},
 \nonumber\\
& \phantom{x}& \nonumber\\
\ll {\mbf G}\gg  &=&  {\mbf G}^{(1)}.
\end{eqnarray}

We calculate the matrix coefficients up to a $n=N_0$ and approximate at coefficients $> N_0$ by {\bf A} and {\bf B}.We then write  for a $N \gg N_0$, 

\[ {\mbf G}^{(N)}\ =\ \left[\rule{0mm}{4mm} (w^2 - i\delta){\mbf I}\right]
^{-1} \]

\n and then iterate 

\begin{eqnarray*}
G^{(n)} = \left[\rule{0mm}{4mm} w^2{\mbf I}\ -\ {\mbf A}\ -\ {\mbf B}^\dagger\ {\mbf G}^{(n+1)}\ {\mbf B}\right]^{-1}& \\
 &  \mbox{ for } n > N_0.
\end{eqnarray*}
 The self-energy follows from the  Dyson equation 
\[ {\mbf \Sigma}\ =\ {\mbf g}^{-1}\ -\ {\mbf G}^{-1}. \]

The Green matrix and self-energy matrices are essential inputs for the calculation of the thermal
conductivity and effective currents.

\subsection{Brillouin zone integration for disordered systems}

The need of efficient techniques for Brillouin zone (BZ) integration in solid state physics has been widely appreciated in recent years. Such techniques are of great importance in the numerical calculation of density of state, conductivity, susceptibility, dielectric function etc. The tetrahedron method allows us a very accurate k-space integration for both the spectral functions and conductivities for ordered systems. Recently our group has developed a generalization of this technique for disordered alloys. The spectral functions are now no longer delta functions, but  Lorenzians with a disorder induced non-zero half-widths \cite{smj}. We will use the efficient codes developed by our group to carry out the integrations over the Brillouin zone. We refer the reader to the above referenced paper for details of the calculation.

\section{RESULTS AND DISCUSSION}
\begin{figure}[t]
\vskip 0.5cm
\centering
\includegraphics[width=3.in,height=2.8in]{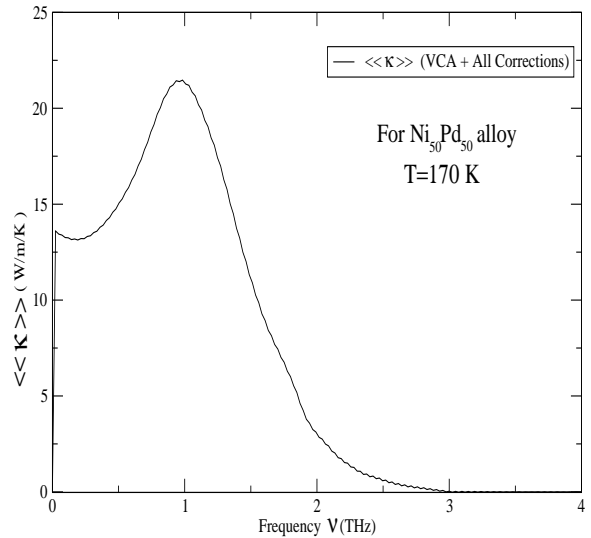}
\caption{Thermal conductivity as a function of frequency for Ni$_{50}$Pd$_{50}$
alloys at T=170 K}
\label{fig12}
\end{figure}

\begin{figure}[b]
\vskip 0.5cm
\centering
\includegraphics[width=3.1in,height=2.8in]{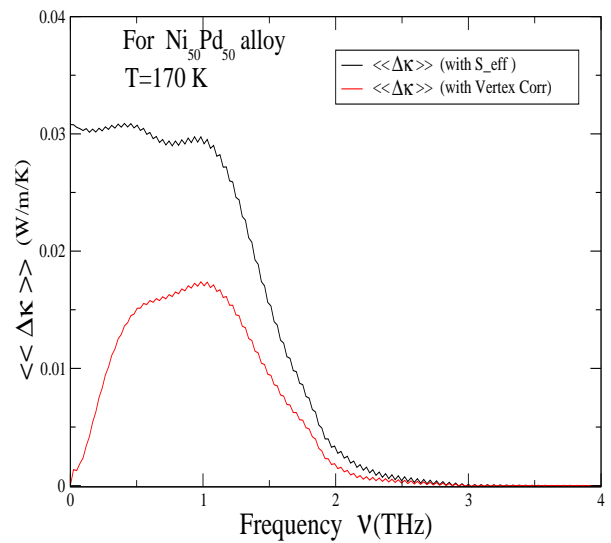}
\caption{(Color Online) Corrections arising due to the disorder renormalization
of the current term and the ladder approximation to the vertex correction}
\label{fig13}
\end{figure}

We have calculated the lattice thermal conductivity of NiPd alloys. Fig \ref{fig12} shows the thermal conductivity as a function of the frequency at T=170 K. 
We shall  analyze the contributions of various disorder correction 
terms to the averaged thermal conductivity at this temperature from Fig \ref{fig13}. The figure shows separately the corrections due to disorder renormalization of the
current terms and that due to the vertex correction. We note that the major contribution comes from the zero-th order term involving only the averaged
currents and averaged propagators. The largest correction comes from the term which arises due to the disorder renormalization of the current terms, while
the vertex correction in the ladder approximation also has a small but
non-negligible contribution.


\begin{figure}[h]
\vskip 0.4cm
\centering
\includegraphics[width=3.3in,height=4.5in]{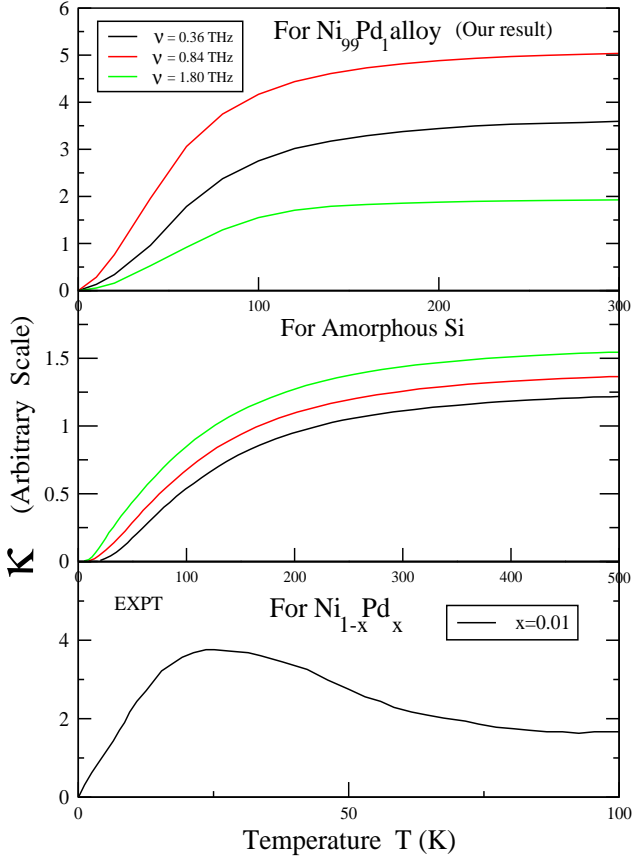}
\caption{(Color Online)\ Thermal conductivity vs temperature T(K) for NiPd alloys and Amorphous Si. The top panel shows our results on the lattice conductivity for Ni$_{99}$Pd$_{01}$ alloy at three different frequency cut-off $\nu$. The middle panel shows the lattice conductivity for amorphous Si \cite{feldmanetal} at three different cut-off frequency, while the panel at the bottom shows the experimental data \cite{expt1} for the total thermal conductivity (=\ lattice + electronic contribution) of the same Ni$_{99}$Pd$_{01}$ alloy. } 
\label{fig14}
\end{figure}

 Direct comparison with the experimental data on these systems is difficult, because the experimental thermal conductivity also has a 
component arising out of the contribution from electrons.
 Figure \ref{fig14} shows  the temperature dependence of lattice conductivity. The top panel shows our theoretical result for the $Ni_{99}Pd_{01}$ alloy at three different frequencies.
The bottom panel shows the experimental data \cite{expt1} on the total ( electronic + lattice  ) thermal conductivity of the same 99-01 NiPd alloy. 
Since the frequency is not mentioned in the
experimental data, we assume that it must be for low frequencies. The best comparison then  will be between the
middle (black) curve on the top panel and that in the bottom one. The two agree qualitatively, except at low temperatures where
we expect the electronic contribution to dominate. In order to understand 
whether the deviation {\sl does arise} from the electronic contribution, we have compared the top panel
with the thermal
conductivity of amorphous-Si  \cite{feldmanetal}, shown in the middle panel. In a-Si the electrons near the Fermi level are  
localized and hence cannot  carry any current. Almost the entire contribution should come from
the phonons. The behaviour of the two panels are quite similar.

\begin{figure}[t]
\vskip 0.5cm
\centering
\includegraphics[width=3.2in,height=3in]{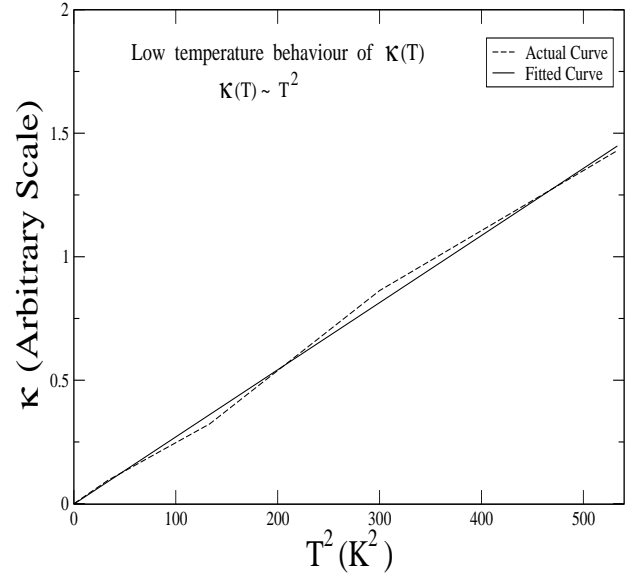}
\caption{$\kappa$ vs $T^2$ for low temperatures for 50-50 NiPd alloy}
\label{fig15}
\end{figure}
 A careful inspection of our results (top panel of Fig. \ref{fig14}) indicates that at low temperatures
 where only low-energy vibrations are excited, $\kappa (T)$ is approximately a quadratic function of $T$. Figure \ref{fig15} shows a plot of $\kappa$ vs $T^2$. The calculated curve fits reasonably well with a straight line.
This has been seen in experimental observations \cite{expt1}.
 Additional scattering processes leading to a different temperature dependence of lattice conductivity become apparent at higher temperatures. At $T>$25 K, $\kappa (T)$ rises smoothly to a $T$-independent saturated value.  
 The dominant mechanism in this regime is the intrinsic harmonic diffusion of higher energy delocalized vibrations.
 These modes have not been well described by most previous theories.

The middle panel for amorphous Si has the similar qualitative behaviour for $\kappa (T)$ as ours. The three curves in this panel stands for three different frequencies. It is clear from the two panels (top and middle) that the saturation of $\kappa (T)$ for NiPd alloy starts at an earlier temperature
 as compared to that for amorphous Si. The magnitude of lattice conductivity in NiPd alloy is 
also larger than the value in amorphous Si, which is obvious because NiPd alloy is a metallic system with larger heat conduction.

In the bottom panel, which shows the experimental data for total thermal conductivity $\kappa=\kappa_e +\kappa_g$, where $\kappa_e$ is the electronic contribution and $\kappa_g$ the lattice contribution to the thermal conductivity, one can notice various similarities with the results of top panel. For example both the panels have (i) an approximate $T^2$-dependence \cite{expt1} in the low temperature regime and (ii) the value of $\kappa(T)$ approaching  a T-independent final or saturated value at higher temperature side. In spite of these similarities, the overall agreement between the top and bottom panels is not satisfactory. This is mainly due to the existence of a broad peak like behaviour in the experimental curve unlike the theoretical results. This feature has been described in  detail for a series of dilute alloys like Ni-Co, Ni-Fe and Ni-Cu in \cite{expt1}. We would like to ascribe this difference
 between the theoretical and experimental data to the extra electronic contribution to thermal conductivity hidden in the experimental curve. 

\begin{figure}[t]
\vskip 0.2cm
\centering
\includegraphics[width=3.3in,height=3.in]{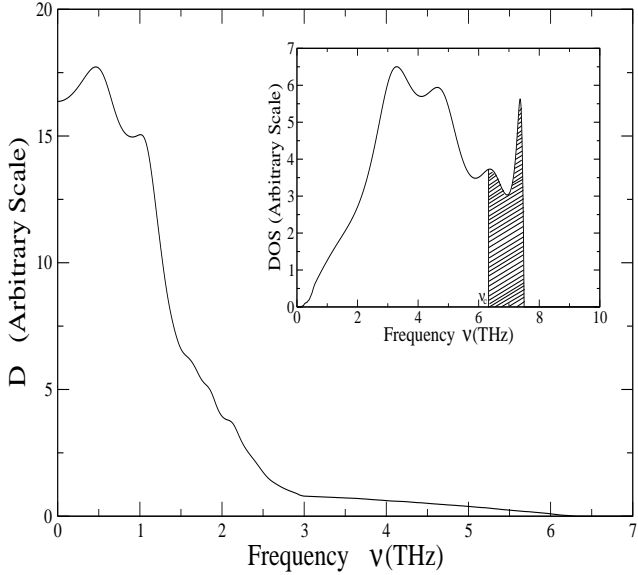}
\caption{The configuration averaged thermal diffusivity $D(\nu)$ vs frequency $\nu$ for Ni$_{50}$Pd$_{50}$ alloy. The inside panel shows the phonon density of states for the same alloy.}
\label{fig16}
\end{figure} 

\begin{figure}[b]
\vskip 0.7cm
\centering
\includegraphics[width=3.3in,height=3in]{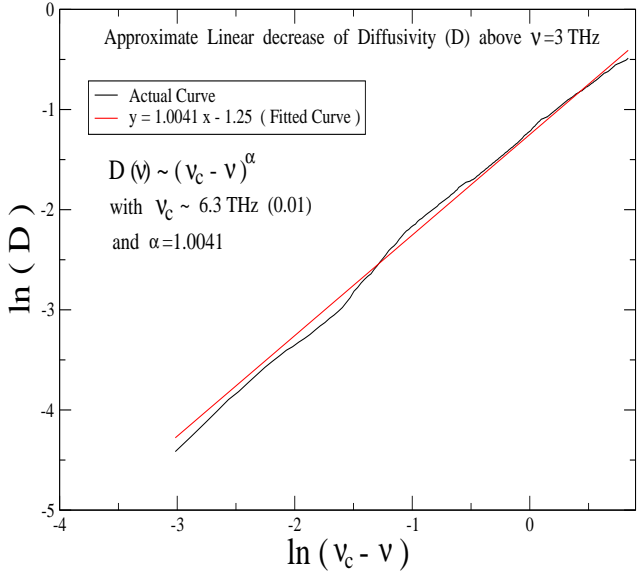}
\caption{(Color Online)\ The behaviour of thermal diffusivity as a function of $(\nu_c-\nu)^\alpha$}
\label{fig17}
\end{figure} 
The thermal diffusivities D($\nu$) are important because the effect of disorder is often manifested in them more directly than in the conductivities. Not only that the thermal diffusivity also give an approximate idea about the location of mobility edge as well as the fraction of delocalized states. Figure \ref{fig16} shows the frequency dependence of diffusivity for Ni$_{50}$Pd$_{50}$ alloy. There are basically two regions of large thermal diffusivity, one near the lower frequency region ($\simeq 0.5 THz$) and the other around a somewhat higher frequency region ($\simeq 1.25 THz$). Above 3 THz, there is a smooth decrease of diffusivity approximately linear in frequency. The Fig \ref{fig17} shows a plot of D($\nu$)$\propto (\nu_c-\nu)^\alpha$. 
 At the critical frequency $\nu_{c}$= 6.3$\pm$ 0.014,  $D(\nu)$ vanishes  to within a very small level of noise. The critical exponent $\alpha \simeq 1.0041$. 
 A knowledge of these critical parameters leads to the information about the location of mobility edge. 
The critical exponent of $\alpha \simeq 1$ in our case agrees with scaling and other theories of Anderson localization \cite{anderson} and the critical frequency $\nu_{c}$ locates the mobility edge above which the diffusivity is zero in the infinite size limit. 
Once the mobility edge is located, the fraction of de-localized states may be obtained by evaluating the area under the $D(\nu)$ vs $\nu$ curve from $\nu=0$ to $\nu=\nu_{c}$. For Ni$_{50}$Pd$_{50}$ alloy (\ see Fig. \ref{fig16}\ ), the mobility edge is located at $\nu_{c}\simeq 6.314 THz$. The panel at the top right corner of Fig. \ref{fig16} shows the phonon density of states for the same alloy. The shaded region in this panel gives the information about the localized states. The fraction of localized states is actually the area beyond the critical frequency $\nu_c$.

\section{Conclusions}
We have formulated a theory for lattice thermal conductivity based on a realistic model. The augmented space approach allows us to go beyond the standard single-site mean-field
theories and the augmented space recursion allows us to include, in the calculation of our averaged propagators, effects of joint fluctuations at more than one site.
The augmented space approach has earlier been generalized to include effects of short-ranged
ordering \cite{sro1,sro2} as well as local lattice relaxations due to large size mismatch of the
constituents \cite{mis}. Calculations including these effects have been earlier carried
out for the electronic case. We propose to apply the same technique to phonons in disordered
alloys. 
 The scattering diagram approach proves to be useful in analyzing and calculating 
 the disorder corrections to the averaged current. These are shown to be  the dominant corrections and are related to the self-energy. 
Next in importance, we have studied the effect of vertex corrections arising out of the correlated propagation. 
We have shown explicitly how to obtain these corrections within the ladder diagram approximation. 
 Our formalism explicitly takes into account fluctuations in masses, force constants and heat currents between different nuclei.
For the calculation of the averaged propagators themselves we have used the augmented
space recursion with the Beer-Pettifor terminator scheme. 
 Our efficient Brillouin zone integration codes for disordered alloys makes the numerical calculation stable and accurate. 
We have already shown in an earlier communication \cite{am2} that the approximation 
involving termination of the matrix continued fraction expansion of the green matrix
 retains the essential herglotz analytic properties of the diagonal green function.
 Our numerical results on the temperature dependence of lattice conductivity favours
 a general trend of other theoretical results as well as the experimental data.
\vskip 0.5cm
\section{\bf Acknowledgments} One of the authors (A.A.) would like to thank the Council of Scientific and Industrial Research, Govt. of India, for a fellowship during the time when this work was carried out.


\begin{thebibliography}{99}
\bibitem{am}A. Mookerjee, J. Phys. C: Solid State Phys. {\bf 6}, L205 (1973).

\bibitem{elliot} R.J. Elliot, J.A. Krumhansl, and P.L. Leath, Rev. Mod. Phys. {\bf 46}, 465 (1974).
\bibitem{kambrock} W.A. Kamitakahara and B.N. Brockhouse, \PR B {\bf 10}, 1200 (1974).
\bibitem{maradudin} A.A. Maradudin, {\it Solid State Physics} Vol. 18, edited by F. Seitz and D. Turnbull (Academic, New York, 1966).
\bibitem{klemens} P.G. Klemes, Proc. R. Soc. A {\bf 208}, 108 (1951)\ ;\ Phys. Rev. {\bf 119}, 507 (1960).
\bibitem{ziman} M.J. Ziman, Can. J. Phys. {\bf 34}, 1256 (1956).
\bibitem{call} J. Callaway, Phys. Rev. {\bf 113}, 1046 (1959).
\bibitem{parrot} J.E. Parrot, J. Phys. C: Solid State Phys. {\bf 2}, 147 (1969).
\bibitem{FL} J.K. Flicker and P.L. Leath, \PR {B} {\bf 7}, 2296(1973).
\bibitem{Tsunoda} Y. Tsunoda, N. Kunitomi, N. Wakabayashi, R.M. Nicklow, and H.G. Smith, \PR B {\bf 19}, 2876 (1979).
\bibitem{glc} S. Ghosh, P.L. Leath, and Morrel H. Cohen, \PR B {\bf 66}, 214206 (2002).
\bibitem{am1} A.Alam and A. Mookerjee, \PR {B} {\bf 69}, 024205 (2004).
\bibitem{af} P.B. Allen and J.L. Feldman, \PR {B} {\bf 48}, 12 581 (1993).
\bibitem{leath} P.L. Leath, \PR {B} {\bf 2}, 3078 (1970).
\bibitem{am2} A.Alam and A. Mookerjee, \PR {B} {\bf 71}, 094210 (2005). 
\bibitem{sm} K.K. Saha and A. Mookerjee, \PR {B} {\bf 70}, 134205 (2004).
\bibitem{hardy} R.J. Hardy, \PR {\bf 132}, 168 (1963). 
\bibitem{mook2} A. Mookerjee, J. Phys. C: Solid State Phys. {\bf 8}, 1524 (1975).
\bibitem{tf} A. Mookerjee, in {\sl Electronic Structure of Alloys, Surfaces and Clusters}, edited by A. Mookerjee and D.D. Sarma. (Taylor and Francis, London, 2003).
\bibitem{mook3} A. Mookerjee, J. Phys. C: Solid State Phys. {\bf 9}, 1225 (1976).
\bibitem{ed} S.F. Edwards, { Philos. Mag.} {\bf 3}, 1020 (1958).
\bibitem{la} J.S. Langer, \PR {\bf 120}, 714 (1960) ; { J. Math. Phys.} {\bf 2}, 584 (1961).
\bibitem{gdma}S. Ghosh, N. Das, and A. Mookerjee, { Int. J. Mod. Phys.} B {\bf 21}, 723  (1999). 
\bibitem{leath-diag} P. L. Leath, Phys. Rev. {\bf 171}, 725 (1968).
\bibitem{aekl} P.N. Aiyer, R.J. Elliott, J.A. Krummhansl and P.L. Leath, Phys. Rev. {\bf 181}, 1006(1969).
\bibitem{nk} B.G. Nickel and J.A. Krummhansl, \PR {B} {\bf 4}, 4354 (1971).
\bibitem{smj} K.K. Saha, A. Mookerjee and O. Jepsen, \PR {B} {\bf 71}, 094207 (2005)
\bibitem{feldmanetal} J.L. Feldman, M.D. Kluge, P.B. Allen, and F. Wooten, \PR {B} {\bf 48}, 12 589 (1993).
\bibitem{expt1} T. Farrell and D. Greig, J. Phys. C : Solid State Phys. {\bf 2}, 1465 (1969).
\bibitem{anderson} P.A. Lee and T.V. Ramakrishnan, Rev. Mod. Phys. {\bf 57}, 287 (1985)
\bibitem{sro1} A. Mookerjee and R. Prasad, \PR {B} {\bf 48}, 17724 (1993).
\bibitem{sro2} T. Saha, I. Dasgupta and A. Mookerjee, \PR {B} {\bf 50},13267 (1994).
\bibitem{mis} T. Saha and A. Mookerjee, \JPCM{\bf 8}, 2915 (1996).
\end{thebibliography}
\end{document}